# The Metallicities of the Broad Emission Line Regions in the Nitrogen-Loudest Quasars

Neelam Dhanda Batra[1,2*] and Jack A. Baldwin[1]

[1]*Physics & Astronomy Department, Michigan State University, East Lansing, MI 48864-1116, USA*

[2]*Malaviya National Institute of Technology, Jawahar Lal Nehru Marg, Malviya Nagar, Jaipur, Rajasthan 302017 India.*

[*]*neelam.dhanda@gmail.com, baldwin@pa.msu.edu*

## ABSTRACT

We measured the metallicity *Z* in the broad emission line regions (BELRs) of 43 SDSS quasars with the strongest N IV] and N III] emission lines. These N-Loud QSOs have unusually low black hole masses. We used the intensity ratio of N lines to collisionally-excited emission lines of other heavy elements to find metallicities in their BELR regions. We found that 7 of the 8 line-intensity ratios that we employed give roughly consistent metallicities as measured, but that for each individual QSO their differences from the mean of all metallicity measurements depends on the ionization potential of the ions that form the emission lines. After correcting for this effect, the different line-intensity ratios give metallicities that generally agree to within the 0.24 dex uncertainty in the measurements of the line-intensity ratios. The metallicities are very high, with mean log *Z* for the whole sample of 5.5 $Z_{\odot}$ and a maximum of 18 $Z_{\odot}$. Our results argue against the possibility that the strong N lines represent an overabundance only of N but not of all heavy elements. They are compatible with either (1) the BELR gas has been chemically enriched by the general stellar population in the central bulge of the host galaxy but the Locally Optimally-emitting Cloud model used in the analysis needs some fine tuning, or (2) that instead this gas has been enriched by intense star formation on the very local scale of the active nucleus that has resulted in an abundance gradient within the BELR.

## 1. Introduction

The broad emission line regions (BELRs) in many quasars have strong nitrogen emission lines, which has been interpreted to mean that the metallicity *Z* (the abundance ratio by number of all metals with respect to H) is very high in these regions, with typical values $Z \sim 4\text{–}5\ Z_{\odot}$ (e.g. Dietrich et al. 2003; Nagao et al. 2006) and with some cases where $Z > 10\ Z_{\odot}$ (e.g. Baldwin et al. 2003a). Some chemical evolution models (Hamann & Ferland 1993, hereafter HF93) indicate that such high metallicities can (barely) be reached through normal stellar evolution in the cores of giant elliptical galaxies, but other models (Friaca & Terlevich 1998; Romano et al. 2002) cannot achieve this. An alternative picture (Collin & Zahn 1999; Wang et al. 2011) is that the BELR gas has been enriched locally within the central part of the AGN rather than by the general stellar population of the host galaxy.

The standard BELR abundance-determination technique (HF93; Hamman & Ferland 1999; Hamann et al. 2002, hereafter H02) takes advantage of the fact that the CNO process in stars enhances the nitrogen abundance relative to the other metals in a way such that the N/O abundance ratio is proportional to the overall metallicity *Z*. The net strength of collisionally-excited emission lines of heavy elements relative to recombination lines such as Lyα does not carry information about metallicity because these two types of lines measure the cooling and heating rates in the gas, respectively, and these are automatically in balance. However, the collisionally-excited lines of different heavy elements compete with one another to carry the cooling load of the gas, so their relative strengths do depend on the relative abundances of heavy elements.

In many quasar spectra, the only N line strong enough to be measured is N V λ1240, and the strength of this line with respect to C IV λ1549 or He II λ1640 is the typical indicator of high metallicity. However, it has long been known from reverberation measurements that these different quasar emission lines are formed in overlapping, but non-identical parts of the BELR (e.g. Peterson 1988; Peterson & Wandel 1999, 2000; Onken & Peterson 2002; Kollatschny 2003). This emphasizes the fact that a correction for the different ionization levels of N ($N^{+2}$, $N^{+3}$, $N^{+4}$), $He^{++}$ and $C^{+3}$ is needed in order to properly determine the relative abundances from the observed line-intensity ratios. These ionization corrections are usually made by assuming a model for the radial (relative to the ionizing continuum source) and density distribution of the gas in the BELR, and then computing how the observed line-intensity ratios are predicted to change with changing metallicity.

The model normally used for this is the Locally Optimally-emitting Cloud (LOC) model (Baldwin et al. 1995). For simplicity it assumes that the distributions in internal gas density *n* and in the radial distances *r* can be described by power laws. The default power laws, which are the ones used by H02 to calibrate the quasar abundance measurements, are that the covering factor of the BELR gas is proportional to $r^{-1}$ and $n^{-1}$. This satisfactorily reproduces the mean spectrum of all QSOs, but is not likely to exactly describe the situation in any particular QSO. Indeed, the spectra of QSOs with unusually narrow emission lines often show lumpy, complex emission-line profiles; demonstrating a non-uniform distribution of clouds over internal gas density and radial position (Baldwin et al. 1996).

Given these uncertainties in the ionization corrections, it is important to try and use lines from other ionization states of N, in addition to N V. Besides providing a check for errors due to overlooked ionization/excitation effects, this will also check for errors in measuring the heavily blended N V line. The first object checked for such a case was Q0353-383, which had been found by Osmer (1980) to have N III] λ1750 and N IV] λ1486 lines that are far stronger than in typical QSOs. Although they are still much weaker than the strongest BELR lines such as Lyα or C IV λ1549, the N III] and N IV] lines in Q0353-383 are strong enough to be accurately measured, which is not usually the case. Baldwin et al. (2003a) obtained new optical/UV spectra of this luminous QSO, and showed that line-intensity ratios involving all three measurable ionization states of N ($N^{+2}$, $N^{+3}$, $N^{+4}$) all give the same result that N/C ~ 15$(N/C)_{\odot}$ and hence the metallicity $Z$ ~ 15 $Z_{\odot}$ in this particular object. This supported the idea that the high metallicities determined for other QSOs in which just the N V line can be measured are also generally correct.

In a follow-up paper (Dhanda et al. 2007) we studied the validity of the abundance measurements in two additional QSOs which had unusually strong, and therefore accurately measurable, N III] and N IV] lines. The objects were SDSS J 125414.27+024117.5 and SDSS J 154651.75+525313.1. In the first of these QSOs, the lines of all the different observed N ionization states again turned out to imply similar values of $Z$, with $Z$ ~ 10 $Z_{\odot}$. However, in the second case, the different line-intensity ratios indicated different metallicities. This implies that in this later object, the standard LOC model does not correctly describe the structure of the BELR. Thus, two out of three of the "N-Loud" QSOs studied to date give results that support the use of the N V line strength as an abundance indicator, but the third object provides a warning that this line is not 100 per cent trustworthy.

Here we identify a new sample of 43 QSOs which, in addition to strong N V lines, have the very strongest N IV] and N III] lines of all of the objects in a very large SDSS quasar sample. We then use the N-line technique to measure Z independently from each of a number of different line-intensity ratios involving various ionization states of nitrogen. Our direct goal is to make improved measurements of the BELR metallicity in QSOs at the very highest end of the metallicity distribution, and to then compare the distribution of metallicities and the dependence of metallicity on other physical parameters of the QSOs to the predictions of chemical enrichment models. We also address the question raised by Jiang, Fan & Vestergaard (2008) about whether the strong N lines really indicate higher overall metallicity as opposed to just a very high abundance of only N. Much of this work has appeared as a PhD thesis (Dhanda 2010, hereafter D10), and further details of the analysis procedure and tables of intermediate results can be found there.

## 2. The quasar samples

We use the fourth edition of the SDSS Quasar Catalogue (Schneider et al. 2007) as the base sample for our abundance studies. The catalogue contains a total of 77,429 objects and is based on the SDSS fifth data release (SDSS DR5; Adelman-McCarthy et al. 2007). It consists of objects that have luminosities greater than $M_i$ ~ -22.0 (in a cosmology with $H_0$ = 70 km $sec^{-1}$ $Mpc^{-1}$, $\Omega_M$ = 0.3, $\Omega_\Lambda$ = 0.7) and have at least one broad emission line (BEL) with FWHM larger than 1000 km $s^{-1}$ or have interesting/complex absorption features, are fainter than i ~15.0 and have highly reliable redshifts. This catalogue covers an area of about 5740 $deg^2$ at high Galactic latitudes. The quasars' redshifts range from 0.08 to 5.41, with a median value of 1.48 and 70 per cent of all detected quasars have redshifts $z$ < 2.0.

### 2.1 The Intermediate sample

Not all spectra in the full quasar catalogue are useful for this study, so we narrowed down the sample to a usable "Intermediate" sample which includes the N-Loud objects of interest but also a much larger number of comparison quasars with similar quality spectra. The abundance analysis uses emission lines ranging in wavelength, at least, from N V λ1240 to C III] λ1909, with extra wavelength coverage needed at either side to allow for continuum fitting. The SDSS spectral coverage (3800–9200Å) restricts us to use the 8122 catalogue objects with $2.29 \leq z \leq 3.61$, which is about 20 per cent of the full number in the quasar catalogue. The individual spectra had already been flux and wavelength calibrated using the standard SDSS pipeline before we downloaded them from the SDSS site. The SDSS Quasar Catalogue includes for each object, the redshift $z$ and the Galactic extinction $A_u$ based on the maps of Schlegel, Finkbeiner & Davis (1998). We used these values with standard IRAF routines to deredden each spectrum, convert it to rest wavelength, apply a 7-pixel boxcar smoothing. We then used the IRAF routine *sfit* to measure and subtract a continuum fitted as a polynomial with maximum order 3, after rejecting outlying pixels caused by atmospheric emission lines, residual cosmic rays etc. This was done for all of the 8122 quasars. Next, we wrote a Fortran program, which automatically identified QSOs with strong C IV λ1549 absorption by comparing the fraction of flux above and below the fitted continuum in a window around the C IV rest wavelength. Objects with more than 50 per cent of the C IV emission line flux absorbed were rejected. This left us with 7788 quasars, which constitute our Intermediate sample.

Finally, we performed automated measurements on the entire Intermediate sample. We wrote another FORTRAN program which extracted the continuum luminosity at 1450 Å, $L_\lambda(1450)$ from the fitted continuum spectra, and which measured the CIV integrated emission-line flux and the full width at half maximum intensity, FWHM (C IV), from the continuum subtracted spectra. The line width was determined by first finding the peak intensity and then searching in both directions in wavelength until the half intensity points were reached.

## 2.2 The N-Loud sample

We next winnowed the Intermediate sample down to a final sample of 43 N-Loud quasars on which we carried out a detailed abundance analysis. To get down to this number we executed the following steps.

First, all of the Intermediate sample spectra were examined by eye and ~2000 of them which were too noisy for detailed abundance analysis were discarded. The S/N threshold for this rejection was roughly 5. We also rejected any spectra which had data missing in the regions of emission lines that were needed for further processing.

We next checked the authenticity of the N III] and N IV] lines, which even when unusually strong, are still weak features blended with the broad wings of C IV (for N IV]) and Fe II (for N III]). Our test was to see if the N IV] and N III] lines had roughly the same profile as the C IV line. This was done by automatically measuring fluxes in the N IV], C IV and N III] lines through a wide velocity window (approximately 10,000 km $s^{-1}$) and then again through a narrower velocity window (approximately 7000 km $s^{-1}$). The N IV]/C IV and N III]/C IV flux ratios were computed separately for each velocity window size, and then the ratio of ratios (narrow/broad) was computed. We discarded cases where this ratio of ratios fell outside the range 0.7-1.3. In the rejected spectra, the apparent N-lines were either just noise, or were dominated by blending with other features.

The rejection steps described above reduced our sample base to about 1000 spectra. We then ordered the surviving spectra by their N IV]/C IV line-intensity ratio. Of these, 42 quasars had N IV]/C IV ratios stronger than in SDSS J125414.27+024117.5, which our previous study (Dhanda et al. 2007) found to have a metallicity of $Z \sim 10\ Z_{\odot}$. An example of a quasar spectrum with strong N lines is shown in Figure 1.

Our present analysis also includes the quasar SDSS J125414.27+024117.5, although with $z$ = 1.84, it lies slightly outside our redshift range. Thus our final sample consists of 43 quasar spectra in all. Final data reduction and full abundance analysis was performed on these 43 quasars, which have the very strongest N lines in the entire Intermediate sample. Table 1 lists these quasars and some of their directly measured properties as well as the black hole mass which is computed as described in the next section. Table 1 gives the SDSS quasar names in full, but in the text and later tables we will use shortened versions of the names that still uniquely identify each object.

## 3. Comparison of Sample Demographics

### 3.1 Redshifts, continuum luminosities and line widths

Figure 2(a) shows that the redshift distributions of the two sample sets are very similar (except for the added object J1254+0241 which is outside the main redshift range). The continuum luminosity distributions are also roughly the same, with the N-Loud sample having mean $L_{\lambda}$(1450 Å) 0.1 dex larger than the Intermediate sample (Figure 2(b)). The largest difference in the directly observed properties, besides the strength of the N lines, is that the N-Loud sample has considerably narrower C IV lines than the Intermediate sample (Figure 2(c)).

### 3.2 Black hole masses

The mass $M_{SMBH}$ of the supermassive black hole at the centre of each quasar (in units of $M_{sun}$) was estimated using equation (7) from Vestergaard & Peterson (2006), but substituting $\lambda L_{\lambda}$(1450 Å) for $\lambda L_{\lambda}$(1350 Å),

$$\log M_{SMBH} = \log\left\{\left[\frac{FWHM(C\ IV)}{1000\ \mathrm{km\ s^{-1}}}\right]^2 \left[\frac{\lambda L_{\lambda}(1450\ \text{Å})}{10^{44}\mathrm{ergs\ s^{-1}}}\right]^{0.53}\right\} + 6.66\ . \quad (1)$$

This equation applies the virial theorem, assuming that FWHM (C IV) is a virial velocity and that the characteristic distance $R_{BELR}$ to the C IV emitting gas can be estimated from the observed relation between $R_{BELR}$ for a particular line and the continuum luminosity. We derived $M_{SMBH}$ for each of the QSOs in both the Intermediate and the N-Loud sample. The values for the N-Loud sample are listed in the last column of Table 1.

Figure 2(d) compares the resulting black-hole mass distributions and shows that the N-Loud sample has systematically lower $M_{SMBH}$ than the general sample of quasars. We verified this using the Student's $t$ test. We found $t \sim 5.3$, corresponding to a probability $P \leq 0.0001$ that the two samples are drawn from the same distribution. No evolution is seen in the mass of the supermassive black hole with redshift, for either of the two samples (see figure A.13 of D10).

Since the black hole masses are calculated from continuum luminosities and line widths, it is important to check the correlations between these quantities to understand the origin of the difference between the mass distributions for the two samples. Figure 3 shows that both samples have similar noisy correlations between $M_{SMBH}$ and the continuum luminosity. The correlation is highly significant for both samples, and the regression lines (shown on the figure) show that the N-Loud sample systematically falls either roughly 0.4 dex in mass below the Intermediate sample, or equivalently 0.5 dex towards higher luminosities.

$M_{SMBH}$ shows a much cleaner correlation with *FWHM (C IV)* in both the samples (Figure 4), and here it is clear that the large difference in the $M_{SMBH}$ distributions for the two samples is largely due to the difference in their FWHM distributions. Given the relative weakness of the N lines, this might in principle be due to a selection effect where narrower N lines are easier to detect. However, we identified the N-Loud objects using an automated procedure which our tests show are insensitive to such effects for lines-widths at least up through log *FWHM (C IV)* = 3.5, which are essentially the broadest lines seen in either the Intermediate or the N-Loud samples. We therefore conclude that there is likely a real difference, with the N-Loud objects having systematically lower black hole masses than the quasar population in general.

## 4. Abundance Analysis

### 4.1 Fe II subtraction

The next step with the N-Loud sample spectra was to fit and subtract templates of Fe II emission. A grid of such templates based on Vestergaard and Wilkes' (2001) study of I Zw 1, but broadened in velocity by different amounts, was kindly provided to us by M. Vestergaard. These particular templates do not include Fe III. For each N-Loud QSO, we used the template with the broadening which most closely matched the measured widths of strong and relatively unblended emission lines (N IV] or C IV). To the extent possible, the fit was guided by the strength of the "UV bump" in the λλ2240 –2650 Å rest wavelength region, where the Fe II emission normally is the strongest.

A few spectra showed no visible Fe emission, so nothing was subtracted. None of these QSOs for which it was possible to fit Fe II have very strong Fe II bumps, so this correction was modest in all cases. The Fe II strengths relative to the continuum are given in Table 1, in units of $F_{\lambda}$(Fe II)/$F_{\lambda}$(continuum) at $\lambda_{rest}$ = 2450 Å.

### 4.2 Line intensity measurements

We then measured the emission-line strengths by finding and isolating individual lines, which we could use as template velocity profiles for fitting to weak and/or blended lines. This is the same technique that was used by Baldwin et al. (2003a) and Dhanda et al. (2007) to measure the emission line strengths in other QSOs with strong N lines, and examples of the profile fitting are shown in those papers. The template profiles are based on different emission lines for different QSOs, and are listed in table B.2 of D10. The deblending procedure used the template profiles to construct synthetic blends in order to isolate the contributions of the lines that would be used in the abundance analysis. All of the lines used in those synthetic blends are marked on our Figure 1 The measured fluxes of the emission lines are given here in the Appendix (Table A1), with fluxes in units of the total flux in the C IV doublet.

We also measured minimum and maximum fluxes for each line, based on the best fit with alternate template profiles that did not fit as well as the best fitting template. In this or any other technique for measuring the strengths of broad and/or blended QSO emission lines, the errors usually are not due to photon counting or the statistics of the fit, but rather are completely dominated by the uncertainties in systematic effects such as differences between the template profiles and the true shape of the line profile, whether or not additional weak lines should be included in the blend, and errors in estimating the level and shape of the underlying continuum. The maximum and minimum fluxes listed in Table A1 are our best estimates of these effects.

This measurement technique has the advantage that it uses empirically determined line profiles, which in real life exhibit a wide variety of non-Gaussian shapes. It has the disadvantage that different lines in the same QSO can have quite different profiles, so the template profile often does not provide an exact fit. The initial fits were done by eye, and then the exact line strengths were finalized automatically using a $\chi^2$-minimisation technique.

Apart from the lines discussed above for use in the abundance analysis, the O VI $\lambda$1035 line can also be employed for this purpose. We were able to measure the O VI line in 19 of the 43 quasar spectra in the final sample. This provided us with two additional abundance sensitive line-intensity ratios, N V/O VI and N V/(C IV+O VI) for this subset of the N-Loud sample.

### 4.3 Abundance determinations for the N-Loud sample

To estimate QSO chemical abundances, we used the procedure that was developed by Hamann & Ferland (1993, 1999) and H02. The technique uses the observed line-intensity ratios of nitrogen lines to cooling lines of other elements, and compares them to the ratios predicted by the LOC model of the BELR. Specifically, we used the H02 results for their segmented power-law ionizing continuum shape. As we have previously done for three other N-Loud QSOs (Baldwin et al. 2003a, Dhanda et al. 2007), we adjusted the H02 predictions to the revised solar C, O, and Fe abundances found by Allende Prieto, Lambert & Asplund (2001, 2002) and Holweger (2001). This modification means that the same N V strength (relative to other strong lines) now occurs at about 30 per cent lower Z than before, and was accounted for by subtracting 0.11 from $\log(Z/Z_{\odot})$ values given by H02[3].

Line-intensity ratios and their uncertainties were computed from the measurements in Table A1, and are tabulated by D10. These were then converted to metallicities and their uncertainties for each line-intensity ratio, as described above. Table A2 in the Appendix lists the resulting best metallicity values $\log Z_{ind}$ found for each line-intensity ratio, for each QSO, along with their minimum and maximum values. These results are also shown in Figure 5 as individual plots for each of the 43 QSOs. Further plots showing how the individual line-intensity ratios map into metallicities are shown in figure A.3 of D10[4].

We determined the overall metallicities of each of the 43 QSOs in the N-Loud sample using at least six line-intensity ratios, and for the 19 of them for which O VI] λ1034 could be measured, we used eight line-intensity ratios. From these ratios we computed a mean metallicity $\log Z_{mean}$ (the average of $\log Z_{ind}$), which is listed in Table 2 along with the standard deviation $\sigma(\log Z_{ind})$ of the $\log Z_{ind}$ values used to compute $Z_{mean}$. The $\log Z_{mean}$ value for each QSO is also shown in Figure 5 as the horizontal dotted line in each panel.

### 4.4 The ionization potential vs. metallicity correlation

The metallicities $\log Z_{ind}$ derived from the different line-intensity ratios are correlated with the ionization potentials of the ions which form the lines. Figure 6 shows $\log Z_{ind}$ plotted as a function of $\log Z_{mean}$. Although for many of the $\log Z_{ind}$ the points fall close to the diagonal line representing perfect agreement with $\log Z_{mean}$, there are clear offsets in several cases (N III]/C III], N IV]/C IV, N V/He II, and N V/O VI). For each $\log Z_{ind}$, we computed separate values of $(\log Z_{ind} - \log Z_{mean})$ for each QSO in the N-Loud sample and then averaged them to find the average metallicity offset $<\Delta\log Z_{ind}>$. These results are listed in Table 3, together with the ionization potentials needed to form the ions in the numerator and denominator of the line-intensity ratio, the difference between those ionization potentials, and the average of the ionization potentials. Figure 7 shows that there is a clear correlation between $<\Delta\log Z_{ind}>$ and the average ionization potential involved in each line-intensity ratio, with only $Z_{\mathrm{N\,IV]/C\,IV}}$ lying significantly off the trend. The correlation has $r^2 = 0.58$ if all line-intensity ratios and data points are included. Dropping the N IV]/C IV ratio raises this to $r^2 = 0.84$ if all available QSOs are used for each line-intensity ratio, and $r^2 = 0.85$ if only the 19 QSOs for which O VI can be measured are used. The $<\Delta\log Z_{ind}>$ in Table 3 and Figure 7 were computed using all QSOs and all available line-intensity ratios, but the N IV]/C IV metallicities were excluded when computing the regression line shown in Figure 7 and the fitted values listed in Table 3. If $Z_{mean}$ had instead been computed without using the N IV]/C IV result, it would have been an average of 0.08 dex higher for each QSO and the $<\Delta\log Z_{ind}>$ value would have all been 0.08 dex lower.

[3] Note that the 0.11 correction was inadvertently left out of the metallicities as tabulated by D10, although they have been correctly applied to the x axis of figure A.3 in that paper.

[4] The labels on the tick marks on the y-axes in figure A3 in D10 have their signs reversed, and should run from +1 at the top to -1 at the bottom.

We applied the measured $<\Delta\log Z_{ind}>$ corrections to the individual log $Z_{ind}$ values for each line-intensity ratio in each QSO, by subtracting the values given in the second to last column of Table 3. These new log $Z_{ind,corr}$ values were then averaged together to form a corrected mean log $Z$, which we call simply log $Z_{corr}$ The last two columns of Table 2 list, for each QSO, log $Z_{corr}$ and the standard deviation $\sigma(\log Z_{ind,corr})$ of the log $Z_{ind,corr}$ values used to compute log $Z_{corr}$. Figure 8 shows the values of log $Z_{ind,corr}$ as points and the mean log $Z_{corr}$ as a dotted line for each QSO, in the same format as Figure 5.

## 5. Discussion

### 5.1 The $<\Delta\log Z_{ind}>$ correction

The $<\Delta\log Z_{ind}>$ values correlate well with the ionization potentials of the species which form the emission lines used to compute log $Z_{ind}$. This suggests that these $Z$ corrections have some physical source, rather than just being statistical fluctuations in our small N-Loud sample. Comparison of Figures 5 and 8 shows the effect of applying these corrections: the agreement between the different log $Z_{ind}$ measurements gets worse for some QSOs, but it is better for many others.

To test whether applying the $<\Delta\log Z_{ind}>$ corrections really does represent an overall improvement, we compared the distributions of the scatter $\sigma$ in the log $Z_{ind}$ values, before and after applying the corrections. These are shown in Figure 9, together with the distribution of the error bars on the log $Z_{ind}$ measurements, which is labelled "σ(profile fitting)" in the figure because it reflects the uncertainties in the profile fits used to measure the line intensities. The error bar (or profile fitting) distribution is the histogram of the differences ($\log Z_{ind,best} - \log Z_{ind,min}$) and ($\log Z_{ind,max} - \log Z_{ind,best}$) from Table A.2. Since the error bars are often asymmetrical, each log $Z_{ind}$ measurement contributes two values to the error bar distribution. The upper panel of Figure 9 shows that the typical $\sigma$ values computed without the $<\Delta\log Z_{ind}>$ corrections were significantly larger than the typical uncertainty in the log $Z_{ind}$ due to the uncertainties in the line-strength measurements, while the lower panel shows that after applying the correction the two distributions are in reasonable agreement, with mean values of 0.24 for the $\sigma(\log Z_{ind,corr})$ distribution and 0.21 for the σ(profile fitting) distribution. We conclude that the $<\Delta\log Z_{ind}>$ corrections do produce a real improvement in the internal agreement of the metallicity measurements.

Here we have used the $<\Delta\log Z_{ind}>$ directly as measured from the deviations from log $Z_{mean}$ for all available line-intensity ratios for our small sample of 43 N-Loud QSOs. These are the values listed in the second-to-last column of Table 3. We could instead have used the values from the last column in the table, which are from our fit to the average ionization potential after excluding $Z_{\mathrm{N\,IV]/C\,IV}}$. But we note that both of these sets of corrections are based on the deviations from log $Z_{mean}$ values which include $Z_{\mathrm{N\,IV]/C\,IV}}$, which opens the possibility of recalculating the corrections using mean values which exclude the results from the N IV]/C IV ratio. The application of the $<\Delta\log Z_{ind}>$ does not change the average metallicity of the whole sample. It does however systematically increase log $Z_{corr}$ by 0.04 as compared to log $Z_{mean}$ for QSOs with only six measurements of $Z_{ind}$, and decrease log $Z_{mean}$ by 0.04 for the objects in which O VI could also be measured. This is an artifact of our having averaged the corrections over all 43 QSOs in the sample for the first six $Z_{ind}$ measurements, but used only the 19 QSOs for which O VI could be measured to find $<\Delta\log Z_{ind}>$ for the last two $Z_{ind}$. This effect could be avoided by using only the 19 QSOs to find all of the $<\Delta\log Z_{ind}>$, but we have done that computation and the difference in log $Z_{corr}$ is not meaningful. None of these modifications would make much real difference in the resulting mean log $Z_{corr}$ values, so we have used what seems to us to be the simplest definition of $<\Delta\log Z_{ind}>$.

Previous papers on the metallicity of QSOs have also reported that $Z$ is significantly different as measured using N IV]/C IV (Dietrich et al. 2003; Shemmer & Netzer 2002) and N V/He II (Dietrich et al. 2003) as compared to other line-intensity ratios. Here we have extended that result to $Z_{ind}$ measurements involving additional line-intensity ratios, and shown that there are systematic dependences of the correction factors.

### 5.2 Metallicity and the LOC model of the BELR

The metallicities measured from 7 of the 8 line-intensity ratios studied here usually give reasonable agreement as measured, but the agreement can be significantly improved by including correction terms $<\Delta\log Z_{ind}>$. The initial agreement argues that the LOC model used by H02 to convert from measured line-intensity ratios to metallicity is at least roughly correct on average. But the fact that the $<\Delta\log Z_{ind}>$ values depend on the ionization level of the ions involved suggests that the LOC model could be tweaked to improve the fit.

The LOC model addresses the fact that the observed QSO spectrum shows that the line emission cannot come from just a single gas component with a single ionization parameter, but must instead come from some mix of gas clouds with a range in density and in distance from the ionizing continuum source. Due to the lack of better information, the standard LOC model (Baldwin et al. 1995) assumes that the distributions in radial distance and density can be characterized as simple power laws with an index of -1. Since lines from different ionization levels are produced at different characteristic radial distances, the fact that $<\Delta\log Z_{ind}>$ depends on the ionization potential suggests that the radial distribution of clouds should be modified. An optimization of the relevant power law index would be a first step.

The density distribution in the standard LOC model might also need some adjustment. The $<\Delta\log Z_{ind}>$ value that conspicuously does not fit the ionization potential correlation is $Z_{\mathrm{N\,IV]/C\,IV}}$. The N IV and C IV ions are produced at very nearly the same ionization potential (see Table 3), and therefore form with the same ionization parameter. However, the critical density of the N IV] $\lambda$1486 intercombination line is low enough such that the two lines do not form in quite the same locations in the LOC model (see figure 3d of Korista et al. 1997). Perhaps just modifying the radial distribution would fix this $Z_{ind}$, but the density distribution also has considerable leverage on this line-intensity ratio. It is noticeable that although Figure 7 shows a correlation of $<\Delta\log Z_{ind}>$ with ionization parameter, it is also the case that all of the $Z_{ind}$ involving intercombination lines imply smaller $<\Delta\log Z_{ind}>$ than those using permitted lines. Since the strengths of the intercombination lines are all sensitive to the density distribution, this again suggests that the standard LOC power-law density distribution might need to be modified.

Studies of the velocity dependence of line-intensity ratios in high signal:noise spectra of individual quasars (e.g. Baldwin et al. 1996) show that at least some BELRs have complicated cloud distributions that are not at all like smooth power laws, so the differences between metallicities from different line-intensity ratios for the same quasar are not especially surprising when the LOC model is applied in such cases. There are three QSOs in the N-Loud sample (J0933+0840, J1159+3134 and J1219+0438) which have values of $\sigma(\log Z_{ind,corr})$ which lie on the high side of the measurement error distribution due to the profile fitting (Figure 9, lower panel). Examination of Figure 8 shows that for the last two of these QSOs, almost none of the line-intensity ratios agree with one another in metallicity, so these would be the most obvious cases where the LOC model is not applicable. These three QSOs constitute 7 per cent of the N-Loud sample. The poor fit from the LOC model might be due to especially complex BELR structure or due to excitation by some non-standard continuum shape. The lumps in the BELR structure presumably average out over a large sample of QSOs, which is why the LOC model works as well as it does on the composite spectra which combines many individual QSO spectra together.

However, for 93 per cent of the quasars in the N-Loud sample, the different metallicity indicators are in reasonably good agreement after applying the $\langle\Delta\log Z_{ind}\rangle$ corrections. This indicates that they are useful individually as metallicity indicators, and gives confidence that $Z_{corr}$ is a good measurement of the average metallicity in individual BELRs.

### 5.3 Metallicity distribution

Here we have studied 43 QSOs with the strongest N III] and N IV] lines, suggesting also the highest metallicities. The agreement between abundances measured by different line-intensity ratios strongly supports the interpretation that quasars with strong N lines indeed have a high nitrogen abundance and are super-solar in metallicity. Figure 10 shows the distribution of the deduced $\log Z_{corr}$ values. The mean and median of $\log Z_{corr}$ for the whole sample are 0.74 and 0.71, respectively, with a minimum value of 0.13 and a maximum of 1.25. In linear units, these correspond to mean $Z_{corr} = 5.5\ Z_{\odot}$ and a range from 1.3 $Z_{\odot}$ to 18 $Z_{\odot}$, with 6 of 43 objects (14 per cent) having $Z_{corr} > 10\ Z_{\odot}$.

These results are consistent with our sample representing the upper end of the same $Z$ distributions studied previously by other authors. Dietrich et al. (2003) studied chemical enrichment in 70 high redshift ($z \geq 3.5$) QSOs and found an average metallicity of 5.3 $Z_{\odot}$ with highest metallicities reaching up to 10 $Z_{\odot}$. They found no trend between metallicity and redshift, but a weak positive correlation of $Z$ with continuum luminosity $L$. Nagao et al. (2006), studied 5344 QSOs from SDSS DR2 by forming composite spectra in different redshift ($z$) and continuum luminosity ($L$) bins. They infer the metallicities for different composite spectra based on the LOC model predictions for various line-intensity ratios which are not the same as ours except for N V/C IV and N V/He II. They found no evolution of metallicities with redshift z, in the range $2.0 \leq z \leq 4.5$, which largely coincides with our $2.29 < z < 3.61$ range. The mean metallicities in their redshift bins are in the range 4–5 $Z_{\odot}$. Our mean metallicity is slightly above the upper limits of the mean $Z$ found by Nagao et al. (2006) for their redshift $z$= 2.75 and 3.25 composite spectra, but the lower-metallicity half of our N-Loud quasars extends down to objects with several times lower metallicity than the Nagao et al. (2006) mean values. Nagao et al. (2006) did find a dependence of metallicity on $L$, with the average metallicity to be 5 $Z_\odot$ in low $L$ bins reaching up to 10 $Z_\odot$ in the highest $L$ bins, and our mean $Z_{corr}$ falls in the middle of this range.

### 5.4 Chemical evolution models and extremely metal rich quasars

Our results directly confirm that the N/O and N/C abundance ratios in our N-Loud sample are super-solar. Using the HF93 scaling then indicates overall metallicity $Z$ up to about 20 $Z_\odot$. This is very high, but still consistent with some models of chemical evolution in the bulges of massive QSO host galaxies.

In particular, HF93 computed an important exploratory set of one-zone closed box models which trace the metallicity $Z$ in the gas component of galaxies as a function of time. The shape of the initial mass function (IMF), the star formation rate (SFR) and the timescale for gas infall were varied to produce a range of models describing situations ranging from the solar neighborhood to the bulges of giant elliptical galaxies. The chemical evolution in these models was halted when the remaining gas was less than 3 per cent of the total mass. Their model M4 is appropriate for giant ellipticals, combining a relatively flat IMF (slope $x$ = 1.1, as compared to $x$ = 1.35 for the Salpeter IMF) with a high SFR and a short gas infall timescale of 0.05 Gyr. This model provides a good fit for the observed metallicities in high $z$ QSOs. The metallicity peaks at 10 $Z_\odot$ after 1 Gyr, and then after star formation is halted, falls off to about 6 $Z_\odot$, close to the median metallicity of the N-Loud QSOs studied here. Hamann & Ferland (1999) later showed results from a slightly different giant elliptical model with very similar chemical evolution.

However, this still does not explain the very highest metallicity QSOs found in our study, with $Z_{mean}$ = 15–20 $Z_\odot$. Two other HF93 models explore the effect of putting a 2.5 $M_\odot$ lower limit on the IMF, while retaining the same short gas infall timescale and moving to even higher SFRs. Their model M5 has a flat IMF (slope $x$ = 1) similar to model M4, while their model M6 uses a steeper IMF ($x$ ~1.6). These two models do reach a higher metallicity, $Z$ = 35 $Z_\odot$, although HF93 note that they are rather unphysical owing to absolute lack of low mass stars.

More complex chemical evolution models describing massive galaxies also quickly reach supersolar metallicities, although not quite as high as those indicated by our N-Loud sample. Friaca & Terlevich (1998) used multi-zone models that follow the dynamical evolution of the gas by including the several episodes of gas inflows and outflows and in particular the evolution of a galactic wind. They adopted the Saltpeter IMF with slope $x$ = 1.35. In the example for which they describe results, the evolution in the chemical abundances is very similar to HF93 model M3 (which uses a similar IMF), with a maximum value of Z ~ [O/H] ~ 4 $Z_\odot$ reached after 1 Gyr. Romano et al. (2002) also discussed the chemical evolution in massive spheroids at high redshift, using a one-zone ISM and taking into account the effects of cooling and stellar feedback. They found a similar 1 Gyr timescale for reaching the peak metallicity, but the peak metallicity reached ($Z \sim 1.3 Z_\odot$) was much lower than that in the HF93 models.

Our sample of 43 N-Loud QSOs has selected the highest metallicity end of a broader range in $Z$ among the thousands of SDSS QSOs. One possible way to interpret these results is in the context of a proposal by Silk & Rees (1998) that QSOs gestate out of sight, enshrouded by dust in the centres of galaxies, and then only become visible at the very end of the process of building a massive black hole, when the luminosity becomes high enough to blow away or evaporate the remaining dust. This carries with it the implication that in most QSOs we are seeing the final end-state reached in the chemical enrichment process. If so, then the different metallicities measured in the general SDSS sample would indicate that different host galaxies shut off metal enrichment at different metallicities. One explanation for such differences, suggested by the HF93 models, would be that there are differences in the IMFs in the host galaxies in a situation where chemical enrichment is stopped because essentially all of the gas has been turned into stars as is inherent in the HF93 models.

However, an alternate possibility fitting within the same general picture is that the enrichment is in fact halted by feedback from the QSO, and that $L \sim L_{EDD}$ is reached at different points in the enrichment process for different QSOs. A third possibility is that the QSOs are not as enshrouded as was suggested by Silk & Rees (1998), and that in fact we are seeing similar objects, all of which will reach, $Z$~20 $Z_{\odot}$, at different moments in their chemical evolution. This could be tested by converting the predicted $Z(t)$ curves for each model into histograms of the expected number of QSOs as a function of $Z$, and seeing if any of the models come at all close to fitting the observations. That would require a very careful combining of the distribution of metallicities measured here for our N-Loud sample with the distribution of metallicities measured for lower-metallicity objects by using just the N V/C IV ratio, in order to get the widest possible range in $Z$.

### 5.5 What does the N/C abundance ratio really measure?

Our analysis technique directly measures abundance of N relative to C and O. We then assume that the chemical abundances are set up by a stellar population whose chemical history is such that N is a secondary element while the other metals that are measured (C, O etc.) are primary elements. In the simplified chemical enrichment models used by HF93, this means that N/C $\propto$ N/O $\propto$ (O/H). This is known to be the case in the ISM. In this situation, N lines can selectively become stronger by carrying more of the cooling load as N becomes more abundant out of proportion to the other heavy elements, but the corresponding loss in cooling power carried by lines of other heavy elements is spread out over many elements, so each of the other lines only becomes slightly weaker relative to the ionizing continuum.

Nagao et al. (2006) had previously found that some line-intensity ratios not involving N lines are also sensitive to the metallicity. Specifically, they discussed using the strength of the O IV + Si IV $\lambda$1400 blend relative to C IV, and the Al III] $\lambda$1857/ C IV ratio. The Al III] line comes from a lower ionization region on the log $n$ − log $r$ plane than the C IV line, so their ratio is quite model dependent, as can be seen in Nagao et al. (2006) figure 36. However, the O IV and Si IV lines do come from gas with about the same ionization parameter as C IV, so they should be a useful abundance indicator. Nagao et al. (2006) found that the (S IV + O IV)/C IV intensity ratio should change by about a factor of five as the metallicity $Z$ is changed by a factor of 100 following the same metal enhancement prescription that was used in the H02 models. The N V/ C IV ratio changed by a factor 50 for the same metallicity change. A large part of the difference in metallicity dependence is because N is a secondary element, with N/O $\propto$ O/H, while O, C and Si are all primary elements. In practical terms, this means that given all the other things that change in the BELR from object to object, the metallicity signature in the (S IV + O IV)/C IV ratio will be far harder to detect than that in the N V/C IV or other similar line-intensity ratios comparing N to primary elements.

Jiang et al. (2008) studied an N-Loud sample of 293 QSOs and suggested that while the N lines are strong, the lines of other elements showed fairly normal ratios so that perhaps only the N abundance is enhanced. Our discussion above suggests that this is not a strong argument. About half of the objects in our N-Loud sample (24 of 43 objects) are also in the Jiang et al. (2008) sample, which means that about 8 per cent of the Jiang et al. (2008) objects are in our sample. Tables 4 and 5 compare properties of objects in both samples to objects in just one of the samples. We used values tabulated by Jiang et al. (2008) for Table 4, and our own values to generate Table 5. The objects that are in both samples have stronger N III] lines and weaker S IV, C IV and C III] lines than those that are only in the Jiang et al. (2008) sample, even though the two subsets have very similar continuum luminosities. Compared to the objects that are in both samples, the objects that are only in our sample have slightly higher mean metallicity and the individual metallicities $Z_{\mathrm{N\ III]/C\ III]}}$ and $Z_{\mathrm{N\ IV]/C\ IV}}$ are clearly higher (meaning that these two line-intensity ratios are on an average larger). These comparisons show that the overall Jiang et al. (2008) sample is diluted by many objects with emission line spectra much more similar to those of normal quasars than is the case for the objects studied here. The equivalent width comparison suggests that in fact there are significant differences in the emission lines from elements other than nitrogen in our N-Loud sample as compared to the broader sample of QSOs.

In the discussion in the preceding section, we assumed that the BELR gas is fed from the ISM of the inner part of the surrounding host galaxy, so that the BELR abundances give information about the chemical history of the bulge population in that galaxy. The mass budget is consistent with this, since the BELR masses of luminous quasars are at least several hundred $M_{\odot}$, and more likely thousands of $M_{\odot}$, consistent with enrichment in a large stellar population (Baldwin et al. 2003b). But Araki et al. (2012) found that the N abundance is not unusually high in the narrow emission line region (NELR) of one QSO that does have strong N lines from its BELR. This implies that perhaps the BELR gas is not simply the raw gas that has fallen in from the surrounding host galaxy, but has been enriched after falling in. We consider that possibility next.

### 5.6 In situ metal production in the AGN central engine?

An alternate explanation for the source of high metallicity in BELRs is that the metals might be produced not by the general stellar population of the host galaxy's central bulge, but rather on more localised scales associated with the AGN central engine (Collin & Zahn 1999). This possibility has recently been studied in some detail by Wang et al. (2011; hereafter W11). In their model, gravitational instabilities cause very rapid star formation to occur in the outer parts of the black hole's accretion disk, and supernovae in this region will then expel super metal-rich gas, which becomes the BELR. A prediction of this model is that the inner region of the BELR should have a higher metallicity than the outer region, due to a strong radial gradient in the SFR. Since the BELR is photoionized by the central engine, this radial metallicity gradient translates to the prediction that higher-ionization species will have higher metallicity than lower-ionization species. This could be detected by comparing metallicities measured from high-ionization lines (HIL) to those measured from low-ionization lines (LIL).

W11 pointed out that there was already observational evidence that $Z_{\text{N III]/C III]}}$ is systematically lower than $Z_{\text{N V/C IV}}$ in composite spectra formed from a large sample of QSOs (Warner, Hamann & Dietrich 2003, 2004). Here we find that the same effect among the individual QSOs in our extreme N-Loud sample. It is the correlation between $\langle \Delta \log Z_{ind} \rangle$ and average ionization potential shown in Figure 7.

These new results are qualitatively similar to those found by Warner et al. (2003). Our study and theirs, both include two metallicity indicators that use only lower-ionization lines, $Z_{\text{N III]/C III]}}$ and $Z_{\text{N III/O III]}}$, and four indicators that use only high-ionization lines, $Z_{\text{N V/He II}}$, $Z_{\text{N V/C IV}}$, $Z_{\text{N V/O VI}}$ and $Z_{\text{N V/(C IV+O VI)}}$. From their figure 10, we find $\log Z_{\text{HIL}} - \log Z_{\text{LIL}} \sim 0.45$, averaged over their bins in log $M_{SMBH}$ centred at 7.8 and 8.6, which span the peak of our $M_{SMBH}$ distribution. The same line-intensity ratios in our study give $\log Z_{\text{HIL}} - \log Z_{\text{LIL}} = 0.33$, in rough agreement with Warner et al. (2003). This supports the W11 model, at least as far as its prediction of a metallicity gradient within the BELR is concerned.

### 5.7 Metallicity and Black Hole Mass

Our comparison of the overall demographics of the Intermediate and N-Loud samples (Section 3) found that the N-Loud sample contains less massive black holes, so it is of interest to look for a dependence of black hole mass on metallicity within the N-Loud sample. Figure 11 shows plots of log $Z_{ind}$ and their mean vs. log $M_{SMBH}$, after correction for the $\langle \Delta \log Z_{ind} \rangle$ metallicity offsets. The figure also shows, for each case, the values of $r^2$ and of the probability $P$ that the correlation is spurious (using the Fisher $F$ statistic). Most of the $Z_{ind,corr}$ and also their mean value $Z_{corr}$ do not correlate with the black hole mass. However, there is a significant ($P = 0.0001$) positive correlation for $Z_{\text{N V/He II}}$. The N V line is badly blended with Lyα, so the $Z_{\text{N V/He II}}$ correlation with $M_{SMBH}$ might in principal be an indirect result of the correlation between $M_{SMBH}$ and line width, but the other metallicities involving the N V line do not show a similar effect. This correlation deserves further study in a future paper.

The difference in computed black hole mass between the Intermediate and N-Loud samples is due to the difference in line widths, and the lower panel of Figure 4 shows that this effect also occurs within the N-Loud sample. We checked that there is no correlation between the C IV line width and metallicity within the N-Loud sample, and we have argued that the correlation between strong N lines and narrow FWHM (C IV) is *not* a selection effect just due to narrow weak lines being easier to detect.

However, the narrow line widths of the N-loud QSOs (Table 1) do raise a caveat for our assertion that these objects are just the high metallicity tail of the general QSO distribution. Almost all of the N-loud objects have FWHM (C IV) < 2000 km $s^{-1}$, which classifies them as narrow-lined type 1 AGN similar to narrow-line Seyfert 1 galaxies (Osterbrock & Pogge 1985). This class of objects shows various anomalies including strong and highly variable soft X-ray excesses (Dasgupta & Rao 2004). The unusual properties of these narrow-line objects are likely due to a combination of viewing angle effects and unusually high Eddington ratios (Peterson 2011).

## 6. Summary and Conclusions

We have picked out the 43 most nitrogen-loud QSOs that we could find in a sample of about 8000 SDSS quasars. Compared to the larger SDSS sample, our N-Loud sample has slightly higher continuum luminosity and significantly narrower C IV emission lines, which using standard assumptions, implies that the N-Loud quasars have less massive central black holes than the general QSO population. Within the N-Loud sample, the computed black hole mass again correlates strongly with FWHM (C IV) and weakly with $L_{1450}$, as expected. It also correlates strongly with $Z_{\mathrm{N\,V/He\,II}}$ and marginally with $Z_{\mathrm{N\,IV]/C\,IV}}$, but not with the other individual metallicity indicators nor with the mean metallicity.

We carried out a detailed abundance analysis on the N-Loud sample using the HF93 technique in which the strengths of nitrogen lines are compared to those of other heavy element cooling lines. By picking out the objects with the very strongest N lines, we were able to reliably measure the strengths of N IV] λ1486 and N III] λ1750 as well as that of the N V λ1240 line, which is normally the only N line that can be measured in a typical QSO spectrum.

For each of the QSOs in the N-Loud sample we measured individual metallicities $Z_{ind}$ from each of the line-intensity ratios (N III]/C III], N III]/O III], N IV]/O III], N IV]/C IV, N V/He II, N V/C IV, N V/O VI and N V/(C IV + O VI)), except that the O VI line strength could only be measured in a subset of 19 objects. We found that the N IV]/C IV line-intensity ratio gives very low metallicities compared to other line-intensity ratios. This confirms earlier work by Dietrich et al. (2003) and Shemmer & Netzer (2002).

The metallicities determined from the other line-intensity ratios do show systematic variations spanning about 0.6 in log $Z$, but which correlate with the ionization potential of the species which form the lines. We worked out correction factors, which we call $\langle \Delta \log Z_{ind} \rangle$, for the metallicity log $Z_{ind}$ determined from each line-intensity ratio. This correction factor for each particular line-intensity ratio is just the difference $Z_{ind}$ - $Z_{mean}$ for each individual QSO and then averaged over all QSOs. Applying these correction factors to the measured log $Z_{ind}$ values improved the agreement between the different metallicities measurements for each individual QSO to the point that the scatter was no worse than the uncertainty in the metallicity measurements due to the uncertainty in measuring the emission-line strengths. Our corrected mean metallicities, log $Z_{corr}$, have an average uncertainty of 0.24.

We suggest that the systematic differences in the measured $Z_{ind}$ point to the need for refinements in the assumed distribution of BELR clouds in the LOC models used to convert the observed line-intensity ratios to the derived metallicities. However, the different line-intensity ratios, especially after applying the $\langle \Delta \log Z_{ind} \rangle$ corrections, give sufficiently consistent metallicities to reinforce the general idea that BELRs can be described by a LOC-type distributed cloud model in which the N/O ratio has been increased by secondary enrichment of N compared to other heavy elements.

We find that our sample includes objects with extremely high metallicities. The mean over log Z for the whole sample corresponds to $Z_{corr} = 5.5\ Z_{\odot}$, with 14 per cent of the sample having $Z_{corr} > 10\ Z_{\odot}$. This is consistent with our sample lying on the extremely metal-rich tail of the broader populations of quasars studied by other authors.

One possibility is that the BELR gas is just the gas that has fallen in from the surrounding host galaxy, and that its chemical enrichment occurred in the stellar population of the central bulge of that galaxy. We compared our results to the predictions of published galactic chemical evolution models. The HF93 models show that flat IMFs and rapid gas infall are needed to reach $Z \sim 10\ Z_{\odot}$, the value reached by 14 per cent of our sample. The highest metallicity that we found is $Z \sim 18\ Z_{\odot}$. The HF93 models require an exceedingly top-heavy IMF to reach such a high metallicity, and other chemical evolution models (Friaca & Terlevich 1998; Romano et. al. 2002) cannot even reach this metallicity. Our results challenge such models.

An alternative is that the chemical enrichment is the result of a massive starburst within the very most central region – the AGN itself. This was originally suggested by Collin & Zahn (1999). The idea has recently been revisited by W11, who developed a detailed model of star formation due to gravitational instabilities in the outer parts of the AGN accretion disk. In their model, the BELR consists of gas ejected by SNe within that disk, and they predict that the inner, more highly-ionized parts of the BELR should have significantly higher metallicity than the outer, less-ionized regions. We compare the metallicities derived from the high-ionization line-intensity ratios (N V/C IV, N V/O VI and N V/(C IV + O VI)) to those found from low ionization line-intensity ratios (N III]/C III], N III]/O III]) and indeed do find the predicted metallicity difference. This confirms earlier results of Warner et al. (2003, 2004).

# Acknowledgements

We thank the referee for helpful comments. We are grateful to the NSF for support of this work through grants AST-0305833 and AST-1006593, and to NASA for support through HST grant AR-10932. NDB carried out this work as part of her PhD thesis at Michigan State University.

IRAF is distributed by the National Optical Astronomy Observatory, which is operated by the Association of Universities for Research in Astronomy (AURA) under cooperative agreement with the National Science Foundation.

Funding for the SDSS and SDSS-II has been provided by the Alfred P. Sloan Foundation, the Participating Institutions, the National Science Foundation, the U.S. Department of Energy, the National Aeronautics and Space Administration, the Japanese Monbukagakusho, the Max Planck Society, and the Higher Education Funding Council for England. The SDSS Web Site is http://www.sdss.org/. The SDSS is managed by the Astrophysical Research Consortium for the Participating Institutions. The Participating Institutions are the American Museum of Natural History, Astrophysical Institute Potsdam, University of Basel, University of Cambridge, Case Western Reserve University, University of Chicago, Drexel University, Fermilab, the Institute for Advanced Study, the Japan Participation Group, Johns Hopkins University, the Joint Institute for Nuclear Astrophysics, the Kavli Institute for Particle Astrophysics and Cosmology, the Korean Scientist Group, the Chinese Academy of Sciences (LAMOST), Los Alamos National Laboratory, the Max-Planck-Institute for Astronomy (MPIA), the Max-Planck-Institute for Astrophysics (MPA), New Mexico State University, Ohio State University, University of Pittsburgh, University of Portsmouth, Princeton University, the United States Naval Observatory, and the University of Washington.

# Appendix A

Here we present our measurements of each line-intensity ratio in each individual QSO in the N-Loud sample. Table A.1 lists the best flux measurement of each line relative to the total flux in the C IV doublet, and the minimum and maximum acceptable values, measured as described in Section 4.2. Table A.2 lists the metallicities determined from each line-intensity ratio in each QSO for the best intensity measurement and also for the minimum and maximum acceptable intensities. The mean metallicity for each object has already been presented in Table 2.

# TABLES

| Table 1. The N-Loud QSO Sample | | | | | | |
|---|---|---|---|---|---|---|
| Name | Redshift z | log ($L_{1450}$) (erg s$^{-1}$ Å$^{-1}$) | FWHM(C IV) (km s$^{-1}$) | $W_\lambda$(C IV) (rest Å) | Fe II Strength | log $M_{SMBH}$ ($M_{sun}$) |
| SDSS J 003815.92+140304.5 | 2.72 | 41.95 | 1720. | 43.1 | 0.57 | 7.72 |
| SDSS J 025505.93+001446.7 | 2.30 | 42.11 | 1440. | 31.5 | - | 7.65 |
| SDSS J 074520.21+415725.4 | 2.86 | 42.30 | 960. | 26.8 | 0.40 | 7.40 |
| SDSS J 075326.12+403038.6 | 2.94 | 42.85 | 1450. | 47.8 | 0.20 | 8.05 |
| SDSS J 080025.10+441723.1 | 3.56 | 42.33 | 1170. | 12.8 | 0.20 | 7.59 |
| SDSS J 084715.16+383110.0 | 3.18 | 42.62 | 1100. | 12.8 | 0.20 | 7.69 |
| SDSS J 085220.46+473458.4 | 2.42 | 41.88 | 1450. | 14.0 | 0.00 | 7.53 |
| SDSS J 085522.87+375425.9 | 2.30 | 42.16 | 2280. | 48.6 | 0.30 | 8.07 |
| SDSS J 093355.72+084043.0 | 2.63 | 42.16 | 2560. | 30.5 | 0.40 | 8.18 |
| SDSS J 095027.35+123335.9 | 3.09 | 41.83 | 930. | 14.9 | 0.66 | 7.12 |
| SDSS J 095334.95+003724.3 | 2.61 | 42.16 | 620. | 32.0 | 0.14 | 6.95 |
| SDSS J 104229.19+381111.2 | 2.63 | 42.61 | 970. | 9.5 | 0.40 | 7.57 |
| SDSS J 104713.39+095711.3 | 2.44 | 41.97 | 1310. | 19.8 | 0.00 | 7.49 |
| SDSS J 104713.16+353115.6 | 2.68 | 42.69 | 1490. | 54.7 | 0.14 | 7.98 |
| SDSS J 105922.31+663806.2 | 3.08 | 41.90 | 900. | 33.2 | - | 7.13 |
| SDSS J 110013.68+030529.8 | 2.32 | 41.59 | 660. | 43.3 | 0.53 | 6.69 |
| SDSS J 112127.96+123816.1 | 3.02 | 41.76 | 1280. | 65.2 | 0.76 | 7.36 |
| SDSS J 115631.40+133714.9 | 3.33 | 42.36 | 1380. | 27.8 | 0.20 | 7.74 |
| SDSS J 115911.52+313427.3 | 3.06 | 42.90 | 760. | 9.2 | - | 7.52 |
| SDSS J 121913.19+043809.1 | 2.74 | 42.51 | 2580. | 24.1 | 0.20 | 8.37 |
| SDSS J 122205.12+034310.3 | 2.59 | 41.78 | 1900. | 86.7 | 0.33 | 7.72 |
| SDSS J 123450.00+375530.3 | 3.14 | 42.32 | 2620. | 13.5 | 0.40 | 8.28 |
| SDSS J 124158.18+123059.3 | 2.97 | 42.11 | 1420. | 18.8 | - | 7.64 |
| SDSS J 125414.27+024117.5 | 1.84 | 42.58 | 1660. | 17.2 | 0.07 | 8.02 |
| SDSS J 130423.24+340438.1 | 2.56 | 42.48 | 760. | 8.7 | 0.04 | 7.29 |
| SDSS J 132827.07+581836.9 | 3.14 | 42.37 | 900. | 50.1 | 0.13 | 7.37 |
| SDSS J 133317.41+641718.0 | 2.82 | 41.92 | 1070. | 26.4 | 0.54 | 7.29 |
| SDSS J 133923.77+632858.4 | 2.56 | 42.35 | 1210. | 43.7 | 0.20 | 7.62 |
| SDSS J 135604.28+471058.7 | 3.38 | 42.44 | 3310. | 24.6 | 0.10 | 8.55 |
| SDSS J 140432.99+072846.9 | 2.87 | 42.37 | 1840. | 13.9 | 0.48 | 8.00 |
| SDSS J 142915.19+343820.3 | 2.35 | 42.36 | 1970. | 37.4 | 0.30 | 8.05 |
| SDSS J 143048.84+481102.7 | 2.50 | 42.13 | 520. | 9.8 | 0.33 | 6.77 |
| SDSS J 144241.74+100533.9 | 2.89 | 42.39 | 2120. | 29.3 | 0.60 | 8.14 |
| SDSS J 144805.84+440806.4 | 3.29 | 42.07 | 1320. | 18.3 | - | 7.55 |
| SDSS J 145615.82+433954.3 | 3.17 | 41.77 | 790. | 55.9 | 0.20 | 6.95 |
| SDSS J 154534.59+511228.9 | 2.45 | 42.07 | 1280. | 21.4 | - | 7.52 |
| SDSS J 155007.07+023607.6 | 2.37 | 41.81 | 720. | 15.7 | 0.29 | 6.89 |
| SDSS J 164148.19+223225.2 | 2.51 | 42.35 | 1280. | 18.8 | 0.39 | 7.67 |
| SDSS J 165023.36+415142.0 | 2.34 | 42.10 | 730. | 16.7 | 0.20 | 7.05 |
| SDSS J 170704.87+644303.2 | 3.16 | 42.58 | 1350. | 32.9 | - | 7.84 |
| SDSS J 171341.05+325045.3 | 2.97 | 41.82 | 1140. | 20.9 | 0.30 | 7.29 |
| SDSS J 233101.64-010604.1 | 3.52 | 41.72 | 790. | 32.5 | - | 6.92 |
| SDSS J 233930.00+003017.3 | 3.05 | 42.52 | 1760. | 22.1 | 0.69 | 8.04 |

| Table 2. Mean Metallicities | | | | | |
|---|---|---|---|---|---|
| SDSS Name | No. of intensity ratios | log $Z_{mean}$ | σ(log $Z_{ind}$) | log $Z_{corr}$ | σ(log $Z_{ind,corr}$) |
| J0038+1403 | 6 | 0.59 | 0.38 | 0.63 | 0.11 |
| J0255+0014 | 6 | 0.67 | 0.16 | 0.71 | 0.32 |
| J0745+4157 | 8 | 0.66 | 0.19 | 0.62 | 0.14 |
| J0753+4030 | 8 | 0.17 | 0.38 | 0.13 | 0.11 |
| J0800+4417 | 6 | 0.76 | 0.37 | 0.81 | 0.11 |
| J0847+3831 | 8 | 0.91 | 0.26 | 0.87 | 0.28 |
| J0852+4734 | 6 | 0.90 | 0.21 | 0.94 | 0.27 |
| J0855+3754 | 6 | 0.29 | 0.62 | 0.33 | 0.41 |
| J0933+0840 | 6 | 0.48 | 0.71 | 0.52 | 0.54 |
| J0950+1233 | 8 | 0.97 | 0.14 | 0.93 | 0.25 |
| J0953+0037 | 6 | 0.34 | 0.22 | 0.39 | 0.20 |
| J1042+3811 | 6 | 0.98 | 0.18 | 1.03 | 0.28 |
| J1047+0957 | 6 | 0.34 | 0.41 | 0.38 | 0.17 |
| J1047+3531 | 6 | 1.20 | 0.48 | 1.25 | 0.29 |
| J1059+6638 | 8 | 0.69 | 0.28 | 0.65 | 0.27 |
| J1100+0305 | 6 | 0.61 | 0.23 | 0.65 | 0.42 |
| J1121+1238 | 8 | 0.85 | 0.48 | 0.81 | 0.29 |
| J1156+1337 | 8 | 0.74 | 0.40 | 0.70 | 0.26 |
| J1159+3134 | 6 | 0.80 | 0.78 | 0.84 | 0.53 |
| J1219+0438 | 6 | 0.60 | 0.77 | 0.64 | 0.51 |
| J1222+0343 | 6 | 0.73 | 0.40 | 0.77 | 0.16 |
| J1234+3755 | 8 | 0.79 | 0.44 | 0.75 | 0.19 |
| J1241+1230 | 8 | 1.11 | 0.52 | 1.07 | 0.26 |
| J1254+0241 | 6 | 0.93 | 0.19 | 0.97 | 0.14 |
| J1304+3404 | 6 | 1.01 | 0.22 | 1.05 | 0.22 |
| J1328+5818 | 8 | 0.43 | 0.40 | 0.39 | 0.15 |
| J1333+6417 | 6 | 0.72 | 0.39 | 0.77 | 0.14 |
| J1339+6328 | 6 | 0.47 | 0.31 | 0.52 | 0.22 |
| J1356+4710 | 8 | 0.97 | 0.60 | 0.93 | 0.33 |
| J1404+0728 | 8 | 1.17 | 0.33 | 1.13 | 0.13 |
| J1429+3438 | 6 | 0.35 | 0.38 | 0.40 | 0.12 |
| J1430+4811 | 6 | 0.90 | 0.11 | 0.94 | 0.26 |
| J1442+1005 | 8 | 0.66 | 0.37 | 0.62 | 0.13 |
| J1448+4408 | 8 | 1.02 | 0.24 | 0.98 | 0.24 |
| J1456+4339 | 8 | 0.67 | 0.27 | 0.63 | 0.23 |
| J1545+5112 | 6 | 0.56 | 0.37 | 0.61 | 0.17 |
| J1550+0236 | 6 | 0.85 | 0.17 | 0.90 | 0.39 |
| J1641+2232 | 6 | 1.01 | 0.28 | 1.06 | 0.19 |
| J1650+4151 | 6 | 0.61 | 0.33 | 0.65 | 0.17 |
| J1707+6443 | 8 | 0.71 | 0.37 | 0.67 | 0.17 |
| J1713+3250 | 8 | 0.73 | 0.25 | 0.69 | 0.07 |
| J2331-0106 | 8 | 0.62 | 0.34 | 0.58 | 0.12 |
| J2339+0030 | 8 | 0.85 | 0.41 | 0.81 | 0.15 |

**Table 3. Average Metallicity Offsets**

| Line-Intensity Ratio | IP (num) (eV) | IP (denom) (eV) | ΔIP (eV) | IP(Avg) (eV) | Measured $\langle \Delta \log Z_{ind} \rangle$ | Fitted $\langle \Delta \log Z_{ind} \rangle$ |
|---|---|---|---|---|---|---|
| N III]/C III] | 29.6 | 24.4 | 5.2 | 27.0 | -0.24 | 0.03 |
| N III]/O III] | 29.6 | 35.1 | -5.5 | 32.4 | -0.07 | -0.14 |
| N IV]/O III] | 47.4 | 35.0 | 12.4 | 41.2 | -0.01 | -0.10 |
| N IV]/C IV | 47.4 | 47.9 | -0.4 | 47.6 | -0.44 | (-0.03) |
| N V/He II | 77.5 | 54.4 | 23.0 | 65.9 | 0.34 | 0.18 |
| N V/C IV | 77.5 | 47.9 | 29.6 | 62.7 | 0.17 | 0.15 |
| N V/O VI | 77.5 | 113.9 | -36.4 | 95.7 | 0.39 | 0.42 |
| N V/(C IV+ O VI) | 77.5 | 80.9 | -3.4 | 79.2 | 0.19 | 0.28 |

**Table 4. Comparison of mean properties of QSOs in overlapping sample to those only in the Jiang et al. (2008) sample**

| Property | In both samples | Only in Jiang et al. (2008) sample |
|---|---|---|
| Number of objects | 24 | 269 |
| Redshift $z$ | 2.7 | 2.2 |
| $i$ mag | 19.1 | 18.8 |
| $M_{2500}$ | -26.2 | -26.0 |
| Contin. Slope α | -0.5 | -0.8 |
| $W_\lambda$(N IV]) (Å) | 4.8 | 4.5 |
| $W_\lambda$(N III]) (Å) | 7.4 | 5.4 |
| $W_\lambda$(Si IV) (Å) | 5.7 | 8.8 |
| $W_\lambda$(C IV) (Å) | 29.0 | 39.8 |
| $W_\lambda$(C III]) (Å) | 16.0 | 23.5 |

**Table 5. Comparison of mean properties of QSOs in overlapping sample to those only in our sample**

| Property* | In both samples | Only in our sample |
|---|---|---|
| Number of objects | 19 | 24 |
| log $Z$ (N III]/C III]) | 0.37 | 0.59 |
| log $Z$ (N III]/O III]) | 0.59 | 0.72 |
| log $Z$ (N IV]/O III]) | 0.67 | 0.76 |
| log $Z$ (N IV]/C IV) | 0.17 | 0.38 |
| log $Z$ (N V/He II]) | 1.03 | 1.11 |
| log $Z$ (N V/C IV) | 0.81 | 0.97 |
| log $Z$ (N V/O VI]) | 1.15 | 1.20 |
| log $Z$ (N V/(C IV+O) VI]) | 0.87 | 1.02 |
| log $Z_{\mathrm{mean}}$ | 0.66 | 0.79 |
| log $Z_{corr}$ | 0.66 | 0.80 |
| $W_\lambda$ (C IV) | 31.0 | 27.0 |
| log $L_{1450}$ | 42.13 | 42.26 |

* Values for $Z_{ind}$ do not include the $\langle \Delta \log Z_{ind} \rangle$ correction.

**Table A1. Measured Line Intensities Relative to C IV λ1549 Doublet**

| | N V | | | N IV] | | | N III] | | | C III] | | | He II | | | O III] | | | O VI | | |
|---|---|---|---|---|---|---|---|---|---|---|---|---|---|---|---|---|---|---|---|---|---|
| SDSS | Best | min | max | Best | min | max | Best | min | max | Best | min | max | Best | min | max | Best | min | max | Best | min | max |
| J0038+1403 | 0.95 | 0.81 | 1.04 | 0.06 | 0.03 | 0.09 | 0.15 | 0.10 | 0.20 | 0.39 | 0.34 | 0.42 | 0.12 | 0.11 | 0.13 | 0.12 | 0.09 | 0.15 | - | - | - |
| J0255+0014 | 0.57 | 0.47 | 0.66 | 0.20 | 0.16 | 0.23 | 0.25 | 0.24 | 0.27 | 0.37 | 0.32 | 0.43 | 0.19 | 0.18 | 0.20 | 0.15 | 0.12 | 0.16 | - | - | - |
| J0745+4157 | 0.75 | 0.67 | 0.86 | 0.08 | 0.07 | 0.15 | 0.15 | 0.13 | 0.20 | 0.21 | 0.16 | 0.25 | 0.21 | 0.20 | 0.23 | 0.11 | 0.09 | 0.11 | 0.65 | 0.59 | 0.72 |
| J0753+4030 | 0.24 | 0.17 | 0.45 | 0.02 | 0.01 | 0.03 | 0.02 | 0.01 | 0.02 | 0.13 | 0.08 | 0.19 | 0.08 | 0.07 | 0.08 | 0.05 | 0.05 | 0.06 | 0.25 | 0.23 | 0.29 |
| J0800+4417 | 1.30 | 1.51 | 1.51 | 0.09 | 0.09 | 0.14 | 0.16 | 0.17 | 0.17 | 0.25 | 0.30 | 0.34 | 0.12 | 0.15 | 0.13 | 0.11 | 0.14 | 0.12 | - | - | - |
| J0847+3831 | 1.46 | 1.23 | 1.79 | 0.13 | 0.11 | 0.16 | 0.32 | 0.27 | 0.40 | 0.29 | 0.27 | 0.35 | 0.15 | 0.14 | 0.15 | 0.10 | 0.10 | 0.15 | 1.72 | 1.55 | 1.83 |
| J0852+4734 | 1.52 | 1.55 | 1.54 | 0.22 | 0.23 | 0.34 | 0.71 | 0.78 | 0.84 | 0.44 | 0.35 | 0.53 | 0.30 | 0.32 | 0.35 | 0.33 | 0.26 | 0.38 | - | - | - |
| J0855+3754 | 0.79 | 0.61 | 1.00 | 0.02 | 0.01 | 0.03 | 0.06 | 0.05 | 0.09 | 0.21 | 0.15 | 0.25 | 0.11 | 0.09 | 0.12 | 0.07 | 0.06 | 0.08 | - | - | - |
| J0933+0840 | 1.71 | 1.12 | 1.97 | 0.03 | 0.02 | 0.03 | 0.18 | 0.15 | 0.22 | 0.33 | 0.23 | 0.42 | 0.19 | 0.17 | 0.20 | 0.12 | 0.10 | 0.14 | - | - | - |
| J0950+1233 | 1.25 | 1.20 | 1.29 | 0.23 | 0.23 | 0.27 | 0.31 | 0.29 | 0.43 | 0.24 | 0.15 | 0.32 | 0.24 | 0.25 | 0.24 | 0.12 | 0.12 | 0.14 | 0.94 | 0.88 | 0.99 |
| J0953+0037 | 0.28 | 0.24 | 0.27 | 0.05 | 0.03 | 0.07 | 0.07 | 0.07 | 0.08 | 0.24 | 0.19 | 0.27 | 0.10 | 0.08 | 0.12 | 0.08 | 0.06 | 0.10 | - | - | - |
| J1042+3811 | 0.95 | 0.67 | 1.24 | 0.23 | 0.23 | 0.24 | 0.30 | 0.31 | 0.32 | 0.19 | 0.14 | 0.23 | 0.17 | 0.18 | 0.18 | 0.10 | 0.11 | 0.13 | - | - | - |
| J1047+0957 | 0.44 | 0.35 | 0.53 | 0.03 | 0.03 | 0.05 | 0.03 | 0.03 | 0.03 | 0.17 | 0.13 | 0.22 | 0.09 | 0.06 | 0.10 | 0.07 | 0.06 | 0.07 | - | - | - |
| J1047+3531 | 1.61 | 1.34 | 1.56 | 0.11 | 0.10 | 0.12 | 0.47 | 0.40 | 0.46 | 0.22 | 0.18 | 0.24 | 0.05 | 0.05 | 0.05 | 0.08 | 0.06 | 0.09 | - | - | - |
| J1059+6638 | 0.55 | 0.52 | 0.60 | 0.16 | 0.13 | 0.16 | 0.09 | 0.08 | 0.09 | 0.22 | 0.18 | 0.27 | 0.14 | 0.14 | 0.14 | 0.09 | 0.09 | 0.10 | 0.38 | 0.38 | 0.38 |
| J1100+0305 | 0.56 | 0.51 | 0.82 | 0.12 | 0.11 | 0.14 | 0.28 | 0.21 | 0.33 | 0.19 | 0.12 | 0.42 | 0.28 | 0.26 | 0.27 | 0.22 | 0.17 | 0.26 | - | - | - |
| J1121+1238 | 0.54 | 0.40 | 0.71 | 0.05 | 0.04 | 0.06 | 0.17 | 0.17 | 0.20 | 0.20 | 0.14 | 0.27 | 0.07 | 0.06 | 0.09 | 0.06 | 0.05 | 0.07 | 0.12 | 0.11 | 0.13 |
| J1156+1337 | 0.63 | 0.43 | 0.76 | 0.06 | 0.06 | 0.06 | 0.22 | 0.19 | 0.23 | 0.21 | 0.17 | 0.24 | 0.15 | 0.15 | 0.15 | 0.15 | 0.14 | 0.15 | 0.19 | 0.11 | 0.19 |
| J1159+3134 | 2.40 | 2.23 | 2.54 | 0.04 | 0.04 | 0.04 | 0.18 | 0.17 | 0.19 | 0.53 | 0.46 | 0.59 | 0.06 | 0.06 | 0.05 | 0.10 | 0.09 | 0.11 | - | - | - |
| J1219+0438 | 1.48 | 0.98 | 1.79 | 0.03 | 0.03 | 0.03 | 0.03 | 0.02 | 0.05 | 0.35 | 0.27 | 0.43 | 0.06 | 0.05 | 0.07 | 0.04 | 0.04 | 0.05 | - | - | - |
| J1222+0343 | 1.04 | 0.98 | 1.18 | 0.08 | 0.07 | 0.10 | 0.18 | 0.17 | 0.21 | 0.51 | 0.32 | 0.63 | 0.11 | 0.08 | 0.13 | 0.08 | 0.07 | 0.11 | - | - | - |
| J1234+3755 | 1.05 | 0.89 | 1.22 | 0.08 | 0.08 | 0.15 | 0.24 | 0.22 | 0.27 | 0.64 | 0.45 | 0.87 | 0.08 | 0.07 | 0.09 | 0.17 | 0.13 | 0.21 | 0.50 | 0.46 | 0.57 |
| J1241+1230 | 1.75 | 1.70 | 1.93 | 0.10 | 0.09 | 0.16 | 0.38 | 0.32 | 0.45 | 0.41 | 0.35 | 0.48 | 0.14 | 0.14 | 0.17 | 0.15 | 0.15 | 0.18 | 0.47 | 0.45 | 0.61 |
| J1254+0241 | 1.50 | 1.44 | 1.44 | 0.23 | 0.23 | 0.25 | 0.40 | 0.37 | 0.39 | 0.42 | 0.42 | 0.42 | 0.23 | 0.16 | 0.21 | 0.16 | 0.13 | 0.15 | - | - | - |
| J1304+3404 | 1.57 | 1.19 | 1.71 | 0.16 | 0.17 | 0.17 | 0.48 | 0.45 | 0.47 | 0.29 | 0.28 | 0.27 | 0.26 | 0.25 | 0.24 | 0.12 | 0.12 | 0.12 | - | - | - |
| J1328+5818 | 0.57 | 0.43 | 0.68 | 0.04 | 0.03 | 0.04 | 0.05 | 0.05 | 0.05 | 0.27 | 0.23 | 0.32 | 0.12 | 0.12 | 0.13 | 0.12 | 0.11 | 0.12 | 0.47 | 0.45 | 0.49 |
| J1333+6417 | 1.25 | 1.04 | 1.63 | 0.06 | 0.05 | 0.07 | 0.28 | 0.25 | 0.30 | 0.50 | 0.45 | 0.52 | 0.14 | 0.14 | 0.14 | 0.11 | 0.09 | 0.13 | - | - | - |
| J1339+6328 | 0.37 | 0.30 | 0.45 | 0.07 | 0.07 | 0.08 | 0.04 | 0.04 | 0.05 | 0.12 | 0.07 | 0.19 | 0.08 | 0.08 | 0.09 | 0.08 | 0.07 | 0.09 | - | - | - |
| J1356+4710 | 1.52 | 0.94 | 1.90 | 0.05 | 0.03 | 0.06 | 0.16 | 0.10 | 0.20 | 0.48 | 0.35 | 0.73 | 0.11 | 0.10 | 0.12 | 0.06 | 0.03 | 0.09 | 0.28 | 0.25 | 0.33 |
| J1404+0728 | 1.68 | 1.23 | 2.05 | 0.18 | 0.19 | 0.21 | 0.28 | 0.29 | 0.28 | 0.27 | 0.16 | 0.33 | 0.17 | 0.16 | 0.17 | 0.08 | 0.08 | 0.09 | 0.39 | 0.32 | 0.40 |
| J1429+3438 | 0.45 | 0.35 | 0.86 | 0.03 | 0.02 | 0.04 | 0.05 | 0.04 | 0.07 | 0.23 | 0.17 | 0.30 | 0.10 | 0.09 | 0.12 | 0.07 | 0.05 | 0.08 | - | - | - |
| J1430+4811 | 1.19 | 1.11 | 1.42 | 0.32 | 0.28 | 0.37 | 0.56 | 0.50 | 0.72 | 0.41 | 0.30 | 0.46 | 0.26 | 0.26 | 0.34 | 0.28 | 0.25 | 0.32 | - | - | - |
| J1442+1005 | 0.66 | 0.48 | 0.83 | 0.05 | 0.04 | 0.07 | 0.12 | 0.09 | 0.12 | 0.28 | 0.25 | 0.34 | 0.07 | 0.05 | 0.09 | 0.09 | 0.09 | 0.13 | 0.49 | 0.48 | 0.46 |
| J1448+4408 | 1.85 | 1.76 | 2.40 | 0.37 | 0.34 | 0.41 | 0.32 | 0.26 | 0.41 | 0.36 | 0.31 | 0.41 | 0.21 | 0.20 | 0.22 | 0.17 | 0.18 | 0.18 | 1.34 | 1.38 | 1.35 |
| J1456+4339 | 0.47 | 0.45 | 0.43 | 0.10 | 0.09 | 0.11 | 0.10 | 0.09 | 0.11 | 0.25 | 0.17 | 0.28 | 0.12 | 0.12 | 0.12 | 0.07 | 0.05 | 0.09 | 0.31 | 0.27 | 0.31 |
| J1545+5112 | 1.16 | 1.15 | 1.15 | 0.08 | 0.07 | 0.09 | 0.13 | 0.13 | 0.14 | 0.35 | 0.30 | 0.46 | 0.19 | 0.21 | 0.18 | 0.17 | 0.19 | 0.17 | - | - | - |
| J1550+0236 | 0.72 | 0.73 | 0.78 | 0.38 | 0.40 | 0.45 | 0.39 | 0.44 | 0.40 | 0.29 | 0.23 | 0.35 | 0.21 | 0.21 | 0.26 | 0.22 | 0.25 | 0.22 | - | - | - |
| J1641+2232 | 1.95 | 1.63 | 2.28 | 0.35 | 0.34 | 0.38 | 0.53 | 0.51 | 0.68 | 0.53 | 0.38 | 0.61 | 0.31 | 0.28 | 0.32 | 0.26 | 0.25 | 0.26 | - | - | - |
| J1650+4151 | 0.57 | 0.46 | 0.80 | 0.06 | 0.05 | 0.12 | 0.24 | 0.21 | 0.26 | 0.52 | 0.38 | 0.50 | 0.11 | 0.11 | 0.13 | 0.09 | 0.08 | 0.12 | - | - | - |
| J1707+6443 | 0.69 | 0.52 | 0.79 | 0.09 | 0.06 | 0.09 | 0.06 | 0.05 | 0.07 | 0.16 | 0.12 | 0.21 | 0.09 | 0.07 | 0.11 | 0.08 | 0.06 | 0.08 | 0.38 | 0.36 | 0.38 |
| J1713+3250 | 0.90 | 0.64 | 1.16 | 0.08 | 0.06 | 0.10 | 0.14 | 0.13 | 0.14 | 0.21 | 0.11 | 0.31 | 0.19 | 0.18 | 0.20 | 0.10 | 0.09 | 0.12 | 0.63 | 0.53 | 0.89 |
| J2331-0106 | 0.48 | 0.37 | 0.68 | 0.05 | 0.05 | 0.05 | 0.10 | 0.09 | 0.12 | 0.25 | 0.19 | 0.23 | 0.10 | 0.10 | 0.10 | 0.09 | 0.08 | 0.08 | 0.23 | 0.22 | 0.24 |
| J2339+0030 | 1.25 | 1.25 | 1.34 | 0.05 | 0.05 | 0.06 | 0.22 | 0.21 | 0.23 | 0.30 | 0.22 | 0.44 | 0.13 | 0.12 | 0.15 | 0.09 | 0.08 | 0.11 | 0.61 | 0.57 | 0.71 |

**Table A2. Metallicity $Z_{ind}$ Determined from Each Line Intensity Ratio for Each N-Loud QSO**

| | N III]/C III] | | | N III]/O III] | | | N IV]/O III] | | | N IV]/C IV | | | N V/He II | | | N V/C IV | | | N V/O VI | | | N V/(C IV+ O VI) | | |
|---|---|---|---|---|---|---|---|---|---|---|---|---|---|---|---|---|---|---|---|---|---|---|---|---|
| SDSS name | Best | min | max | Best | min | max | Best | min | max | Best | min | max | Best | min | max | Best | min | max | Best | min | max | value | min | max |
| J0038+1403 | 0.28 | 0.07 | 0.47 | 0.55 | 0.27 | 0.84 | 0.53 | -0.51 | 0.87 | 0.14 | -0.27 | 0.38 | 1.16 | 1.01 | 1.29 | 0.88 | 0.78 | 0.99 | - | - | - | - | - | - |
| J0255+0014 | 0.50 | 0.36 | 0.75 | 0.66 | 0.55 | 0.87 | 0.97 | 0.77 | 1.24 | 0.66 | 0.52 | 0.76 | 0.58 | 0.39 | 0.83 | 0.64 | 0.46 | 0.79 | - | - | - | - | - | - |
| J0745+4157 | 0.53 | 0.41 | 0.95 | 0.57 | 0.51 | 0.82 | 0.73 | 0.67 | 1.09 | 0.29 | 0.25 | 0.55 | 0.70 | 0.56 | 0.87 | 0.77 | 0.72 | 0.84 | 0.85 | 0.69 | 1.01 | 0.82 | 0.72 | 0.92 |
| J0753+4030 | -0.17 | -0.37 | 0.14 | 0.01 | -0.06 | 0.13 | 0.15 | -0.23 | 0.65 | -0.46 | -0.51 | -0.16 | 0.58 | 0.41 | 1.07 | 0.19 | 0.04 | 0.52 | 0.72 | 0.41 | 1.19 | 0.33 | 0.14 | 0.68 |
| J0800+4417 | 0.49 | 0.20 | 0.65 | 0.60 | 0.40 | 0.73 | 0.76 | 0.55 | 1.04 | 0.35 | 0.16 | 0.68 | 1.30 | 1.14 | 1.50 | 1.09 | 0.92 | 1.46 | - | - | - | - | - | - |
| J0847+3831 | 0.88 | 0.57 | 1.04 | 0.99 | 0.70 | 1.16 | 0.93 | 0.72 | 1.05 | 0.50 | 0.44 | 0.59 | 1.27 | 1.17 | 1.42 | 1.17 | 1.05 | 1.30 | 0.63 | 0.49 | 0.86 | 0.92 | 0.79 | 1.08 |
| J0852+4734 | 1.08 | 0.84 | 1.47 | 0.80 | 0.61 | 1.26 | 0.69 | 0.44 | 1.14 | 0.69 | 0.60 | 0.95 | 0.94 | 0.59 | 1.08 | 1.19 | 0.99 | 1.42 | - | - | - | - | - | - |
| J0855+3754 | 0.17 | 0.02 | 0.49 | 0.41 | 0.27 | 0.62 | -0.29 | -0.52 | 0.46 | -0.46 | -0.56 | -0.23 | 1.13 | 0.93 | 1.35 | 0.80 | 0.64 | 0.97 | - | - | - | - | - | - |
| J0933+0840 | 0.43 | 0.19 | 0.83 | 0.62 | 0.42 | 0.83 | -0.40 | -0.70 | 0.18 | -0.26 | -0.42 | -0.15 | 1.23 | 0.94 | 1.41 | 1.27 | 0.92 | 1.44 | - | - | - | - | - | - |
| J0950+1233 | 0.96 | 0.60 | 1.41 | 0.90 | 0.75 | 1.18 | 1.19 | 1.05 | 1.31 | 0.71 | 0.68 | 0.78 | 0.96 | 0.90 | 1.00 | 1.06 | 0.98 | 1.15 | 0.95 | 0.82 | 1.07 | 1.02 | 0.93 | 1.12 |
| J0953+0037 | 0.16 | 0.02 | 0.39 | 0.38 | 0.18 | 0.65 | 0.64 | -0.30 | 1.01 | 0.07 | -0.31 | 0.32 | 0.54 | 0.27 | 0.90 | 0.28 | 0.09 | 0.38 | - | - | - | - | - | - |
| J1042+3811 | 1.08 | 0.94 | 1.29 | 0.98 | 0.81 | 1.04 | 1.25 | 1.09 | 1.30 | 0.70 | 0.69 | 0.74 | 1.00 | 0.69 | 1.15 | 0.89 | 0.68 | 1.11 | - | - | - | - | - | - |
| J1047+0957 | -0.01 | -0.18 | 0.15 | 0.16 | 0.09 | 0.26 | 0.59 | 0.36 | 0.79 | -0.13 | -0.25 | 0.08 | 0.94 | 0.61 | 1.25 | 0.51 | 0.33 | 0.65 | - | - | - | - | - | - |
| J1047+3531 | 1.22 | 1.03 | 1.39 | 1.48 | 1.14 | 1.73 | 1.02 | 0.82 | 1.28 | 0.43 | 0.33 | 0.52 | 1.86 | 1.76 | 1.88 | 1.23 | 1.04 | 1.29 | - | - | - | - | - | - |
| J1059+6638 | 0.27 | 0.12 | 0.43 | 0.42 | 0.29 | 0.52 | 1.11 | 0.87 | 1.20 | 0.58 | 0.43 | 0.62 | 0.77 | 0.61 | 0.93 | 0.63 | 0.54 | 0.72 | 1.01 | 0.89 | 1.13 | 0.74 | 0.63 | 0.86 |
| J1100+0305 | 1.05 | 0.37 | 1.38 | 0.55 | 0.34 | 0.78 | 0.59 | 0.39 | 0.78 | 0.45 | 0.41 | 0.55 | 0.39 | 0.33 | 0.63 | 0.63 | 0.56 | 0.85 | - | - | - | - | - | - |
| J1121+1238 | 0.66 | 0.46 | 1.05 | 0.99 | 0.82 | 1.29 | 0.77 | 0.61 | 1.03 | 0.05 | -0.06 | 0.22 | 1.13 | 0.84 | 1.40 | 0.62 | 0.42 | 0.78 | 1.73 | 1.44 | 1.98 | 0.86 | 0.61 | 1.06 |
| J1156+1337 | 0.81 | 0.54 | 1.04 | 0.62 | 0.51 | 0.70 | 0.40 | 0.33 | 0.44 | 0.15 | 0.11 | 0.17 | 0.83 | 0.51 | 0.99 | 0.69 | 0.46 | 0.82 | 1.52 | 1.23 | 2.03 | 0.91 | 0.62 | 1.11 |
| J1159+3134 | 0.23 | 0.08 | 0.41 | 0.71 | 0.54 | 0.90 | 0.40 | -0.02 | 0.65 | -0.04 | -0.19 | 0.11 | 1.98 | 1.87 | 2.08 | 1.49 | 1.30 | 1.67 | - | - | - | - | - | - |
| J1219+0438 | -0.28 | -0.61 | 0.05 | 0.38 | 0.13 | 0.63 | 0.75 | 0.52 | 0.84 | -0.16 | -0.33 | -0.08 | 1.72 | 1.42 | 1.93 | 1.18 | 0.85 | 1.37 | - | - | - | - | - | - |
| J1222+0343 | 0.24 | 0.11 | 0.53 | 0.80 | 0.61 | 0.99 | 0.83 | 0.63 | 1.03 | 0.30 | 0.18 | 0.43 | 1.27 | 1.10 | 1.50 | 0.95 | 0.86 | 1.08 | - | - | - | - | - | - |
| J1234+3755 | 0.27 | 0.09 | 0.48 | 0.59 | 0.44 | 0.82 | 0.50 | 0.30 | 0.95 | 0.28 | 0.26 | 0.60 | 1.40 | 1.23 | 1.59 | 0.95 | 0.83 | 1.09 | 1.23 | 1.01 | 1.42 | 1.07 | 0.91 | 1.22 |
| J1241+1230 | 0.64 | 0.45 | 1.04 | 0.93 | 0.81 | 1.23 | 0.69 | 0.50 | 1.04 | 0.35 | 0.20 | 0.59 | 1.55 | 1.36 | 1.69 | 1.45 | 1.34 | 1.52 | 1.74 | 1.51 | 1.88 | 1.54 | 1.40 | 1.62 |
| J1254+0241 | 0.74 | 0.59 | 0.78 | 0.87 | 0.83 | 1.03 | 1.01 | 1.01 | 1.21 | 0.71 | 0.68 | 0.76 | 1.07 | 1.05 | 1.27 | 1.18 | 1.10 | 1.21 | - | - | - | - | - | - |
| J1304+3404 | 1.10 | 0.91 | 1.27 | 1.16 | 0.92 | 1.40 | 0.96 | 0.85 | 1.18 | 0.60 | 0.47 | 0.71 | 1.04 | 0.68 | 1.26 | 1.21 | 0.83 | 1.49 | - | - | - | - | - | - |
| J1328+5818 | -0.01 | -0.10 | 0.09 | 0.12 | 0.08 | 0.17 | 0.24 | -0.13 | 0.37 | -0.06 | -0.17 | 0.00 | 0.91 | 0.63 | 1.03 | 0.64 | 0.48 | 0.74 | 0.89 | 0.62 | 1.06 | 0.71 | 0.52 | 0.86 |
| J1333+6417 | 0.43 | 0.31 | 0.57 | 0.85 | 0.66 | 1.18 | 0.61 | 0.27 | 0.83 | 0.18 | 0.03 | 0.30 | 1.21 | 1.06 | 1.43 | 1.06 | 0.87 | 1.33 | - | - | - | - | - | - |
| J1339+6328 | 0.26 | 0.04 | 0.51 | 0.21 | 0.14 | 0.29 | 0.80 | 0.71 | 0.90 | 0.24 | 0.16 | 0.31 | 0.91 | 0.62 | 1.06 | 0.42 | 0.28 | 0.56 | - | - | - | - | - | - |
| J1356+4710 | 0.20 | -0.12 | 0.47 | 0.89 | 0.50 | 1.51 | 0.79 | 0.30 | 1.21 | 0.09 | -0.17 | 0.21 | 1.42 | 1.15 | 1.61 | 1.19 | 0.86 | 1.37 | 1.82 | 1.39 | 2.06 | 1.39 | 1.05 | 1.56 |
| J1404+0728 | 0.80 | 0.57 | 1.18 | 1.08 | 0.96 | 1.20 | 1.27 | 1.17 | 1.42 | 0.64 | 0.61 | 0.72 | 1.27 | 1.06 | 1.46 | 1.26 | 0.97 | 1.47 | 1.69 | 1.39 | 2.00 | 1.40 | 1.13 | 1.62 |
| J1429+3438 | 0.05 | -0.10 | 0.33 | 0.34 | 0.20 | 0.61 | 0.50 | -0.16 | 0.78 | -0.18 | -0.40 | -0.02 | 0.89 | 0.54 | 1.27 | 0.52 | 0.38 | 0.85 | - | - | - | - | - | - |
| J1430+4811 | 1.00 | 0.85 | 1.30 | 0.76 | 0.63 | 0.96 | 0.89 | 0.79 | 1.04 | 0.81 | 0.77 | 0.85 | 0.91 | 0.61 | 0.99 | 1.04 | 0.99 | 1.15 | - | - | - | - | - | - |
| J1442+1005 | 0.31 | 0.06 | 0.42 | 0.58 | 0.21 | 0.66 | 0.62 | -0.11 | 0.83 | 0.06 | -0.13 | 0.31 | 1.25 | 0.87 | 1.58 | 0.71 | 0.46 | 0.91 | 0.97 | 0.64 | 1.22 | 0.80 | 0.52 | 1.04 |
| J1448+4408 | 0.69 | 0.37 | 1.10 | 0.74 | 0.48 | 0.98 | 1.24 | 1.01 | 1.41 | 0.85 | 0.75 | 1.00 | 1.22 | 1.06 | 1.50 | 1.32 | 1.13 | 1.65 | 0.98 | 0.77 | 1.27 | 1.15 | 0.97 | 1.44 |
| J1456+4339 | 0.31 | 0.13 | 0.53 | 0.62 | 0.38 | 0.81 | 1.04 | 0.78 | 1.29 | 0.40 | 0.26 | 0.50 | 0.76 | 0.60 | 0.85 | 0.54 | 0.45 | 0.58 | 1.03 | 0.92 | 1.15 | 0.67 | 0.56 | 0.72 |
| J1545+5112 | 0.26 | 0.04 | 0.48 | 0.33 | 0.22 | 0.44 | 0.46 | 0.12 | 0.65 | 0.27 | 0.07 | 0.46 | 1.04 | 0.90 | 1.14 | 1.02 | 0.83 | 1.21 | - | - | - | - | - | - |
| J1550+0236 | 0.99 | 0.71 | 1.28 | 0.71 | 0.62 | 0.79 | 1.12 | 0.98 | 1.29 | 0.86 | 0.78 | 1.06 | 0.69 | 0.40 | 0.95 | 0.75 | 0.61 | 0.95 | - | - | - | - | - | - |
| J1641+2232 | 0.65 | 0.52 | 1.00 | 0.88 | 0.83 | 1.17 | 1.10 | 1.06 | 1.25 | 0.83 | 0.83 | 0.88 | 1.18 | 1.06 | 1.34 | 1.44 | 1.36 | 1.56 | - | - | - | - | - | - |
| J1650+4151 | 0.35 | 0.20 | 0.65 | 0.89 | 0.58 | 1.25 | 0.66 | 0.12 | 1.23 | 0.12 | -0.11 | 0.60 | 0.98 | 0.53 | 1.27 | 0.64 | 0.39 | 0.95 | - | - | - | - | - | - |
| J1707+6443 | 0.25 | 0.03 | 0.51 | 0.34 | 0.17 | 0.55 | 0.89 | 0.68 | 1.11 | 0.32 | 0.11 | 0.40 | 1.18 | 0.83 | 1.38 | 0.73 | 0.54 | 0.85 | 1.14 | 0.90 | 1.34 | 0.87 | 0.63 | 1.03 |
| J1713+3250 | 0.50 | 0.30 | 0.95 | 0.57 | 0.48 | 0.63 | 0.76 | 0.61 | 0.88 | 0.31 | 0.19 | 0.40 | 0.92 | 0.60 | 1.08 | 0.86 | 0.70 | 1.02 | 1.00 | 0.53 | 1.26 | 0.93 | 0.64 | 1.12 |
| J2331-0106 | 0.33 | 0.24 | 0.56 | 0.54 | 0.44 | 0.71 | 0.63 | 0.48 | 0.82 | 0.07 | -0.08 | 0.26 | 0.90 | 0.48 | 1.27 | 0.55 | 0.21 | 0.93 | 1.20 | 0.85 | 1.65 | 0.71 | 0.37 | 1.16 |
| J2339+0030 | 0.55 | 0.26 | 0.99 | 0.85 | 0.63 | 1.18 | 0.62 | 0.25 | 0.86 | 0.08 | -0.10 | 0.28 | 1.28 | 1.06 | 1.45 | 1.06 | 0.90 | 1.28 | 1.22 | 0.96 | 1.46 | 1.13 | 0.95 | 1.34 |

# FIGURES

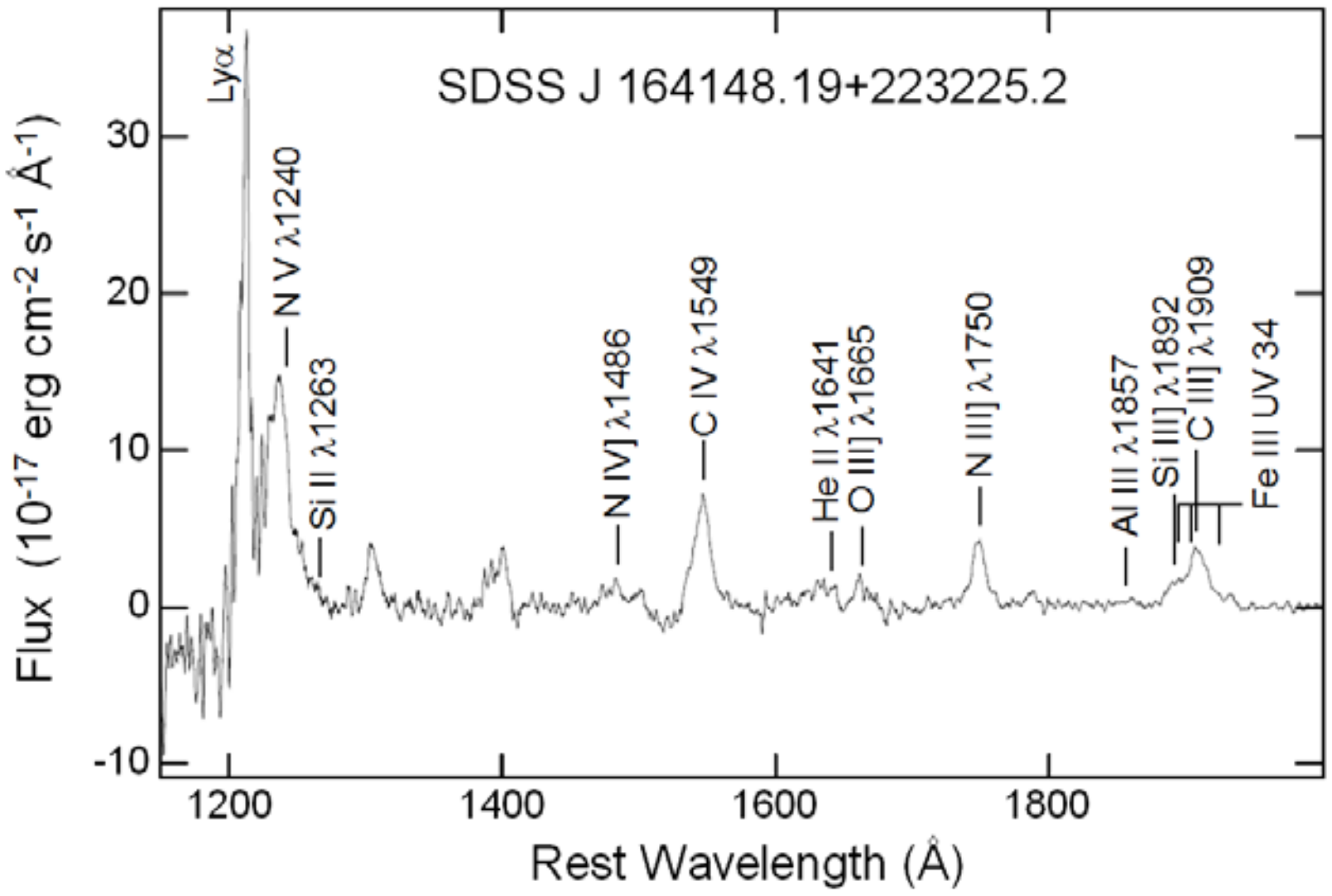

**Figure 1.** Continuum-subtracted spectrum of a QSO from the N-loud sample. All of the metallicity-indicating emission lines and the other lines that were deblended from them are identified, except for O VI λ1034 (which is not within the wavelength range measured for this object) and the very broad Fe II feature.

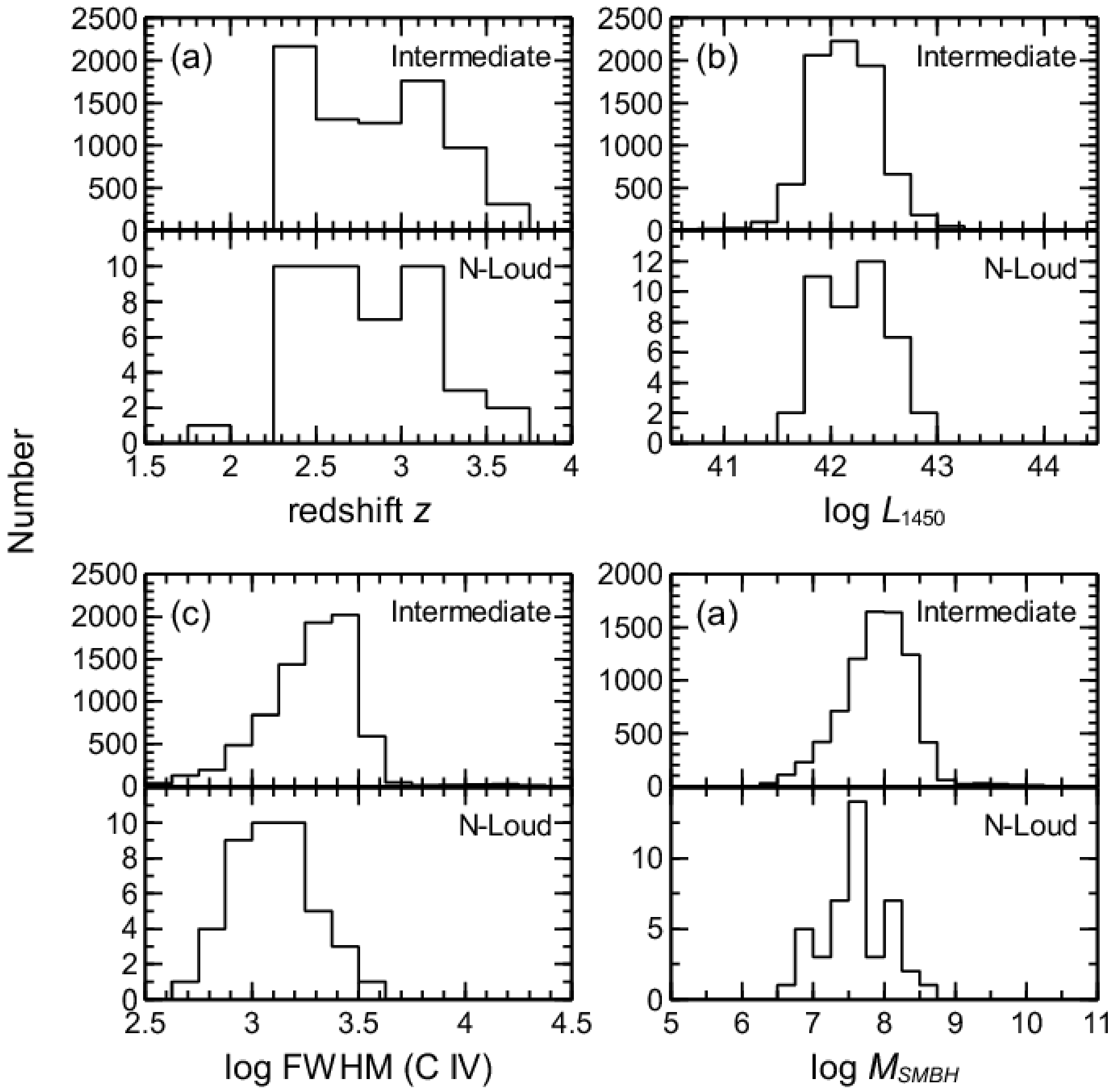

**Figure 2.** Comparison of the properties of the Intermediate and N-Loud samples. The panels show the distributions of (a) the redshift z; (b) the monochromatic continuum luminosity $L_{1450}$; (c) the C IV λ1549 FWHM line width; and (d) the deduced mass of the supermassive black hole, in solar units.

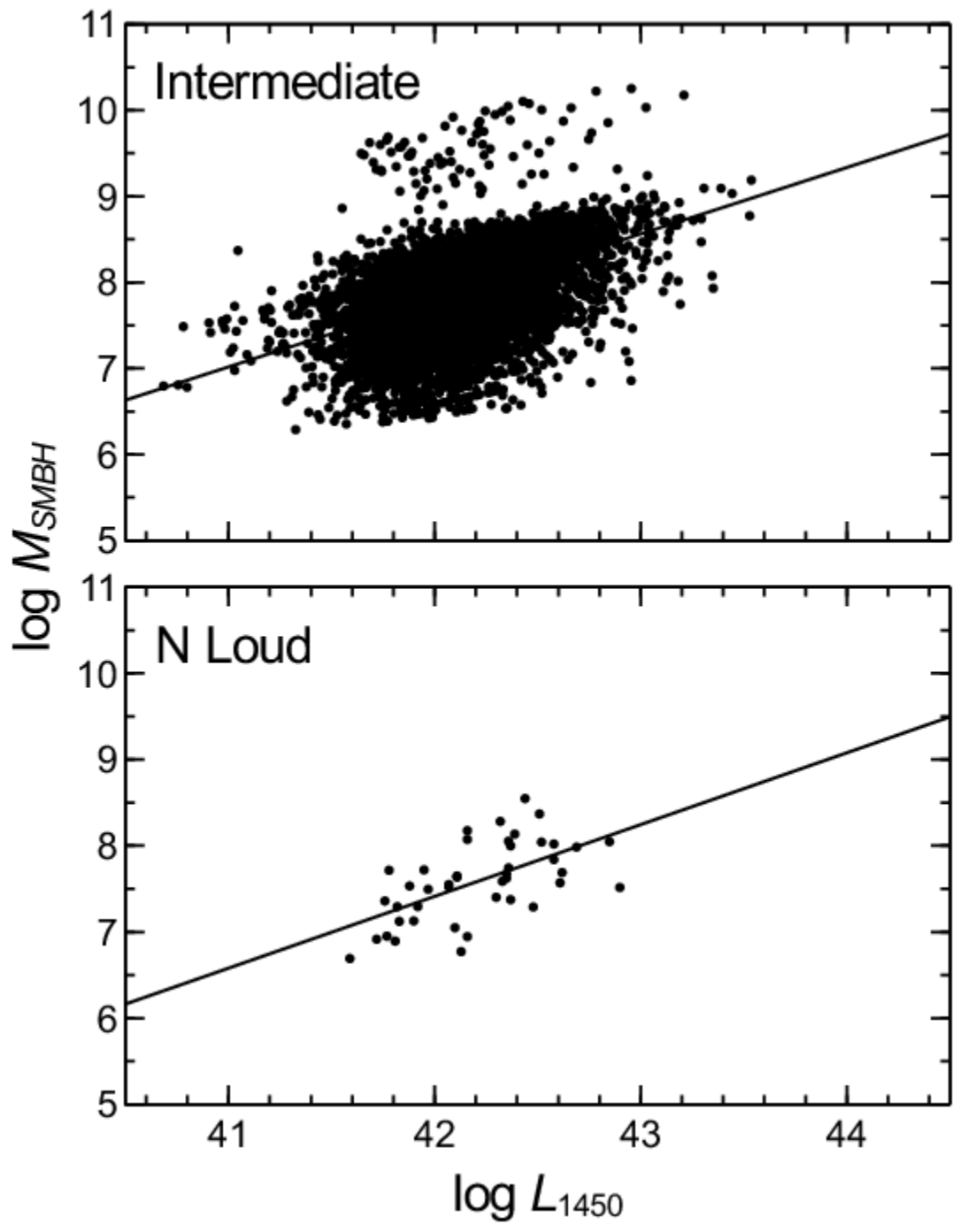

**Figure 3:** Supermassive Black Hole Mass ($M_{SMBH}$) in solar units as a function of continuum luminosity ($L_{\lambda}(1450)$) for the Intermediate sample (upper panel) and the N-Loud sample. The solid lines show the linear regression fits to each sample.

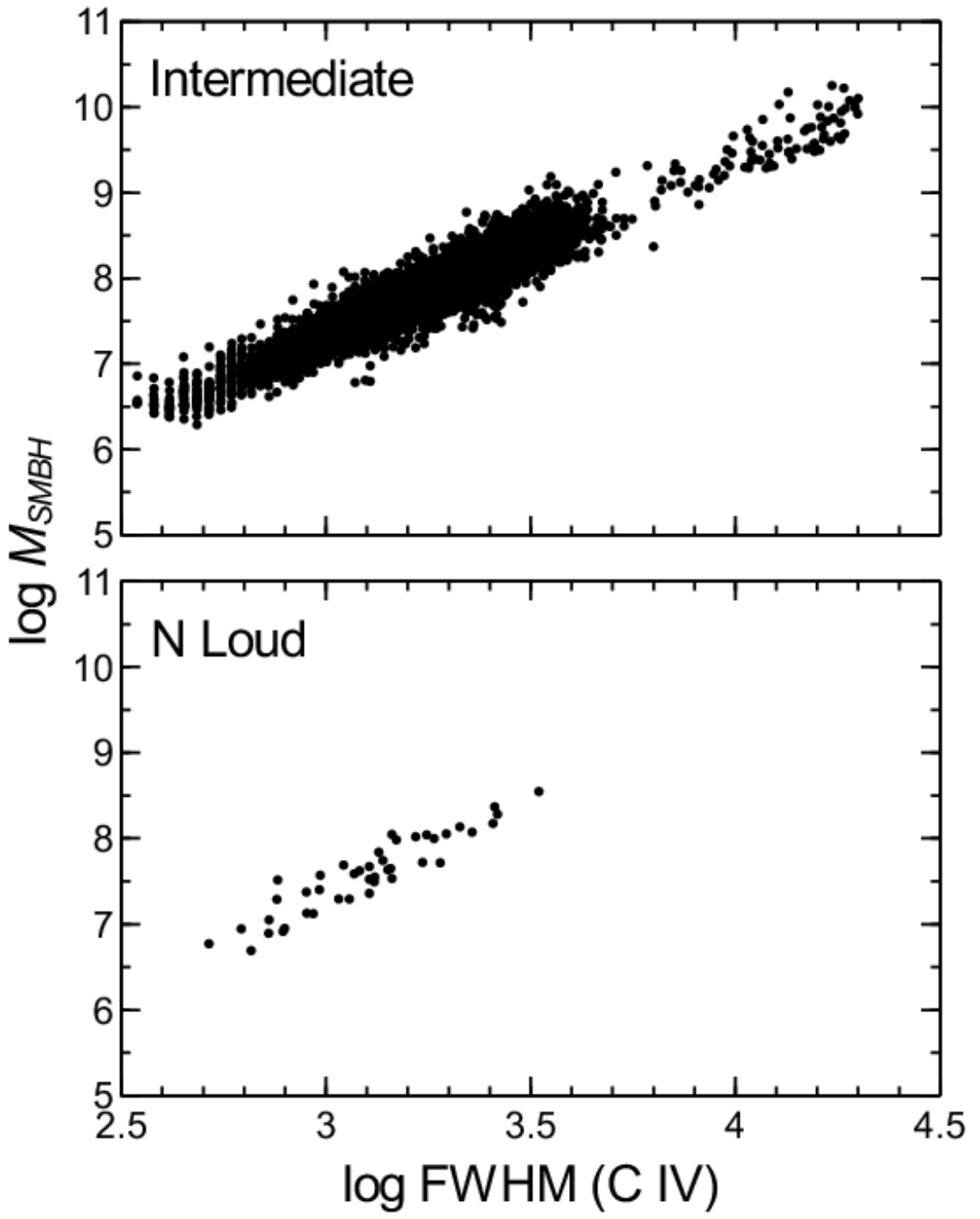

**Figure 4:** Supermassive Black Hole Mass ($M_{SMBH}$) in solar units as a function of FWHM (CIV) in km $s^{-1}$. The top panel shows the Intermediate sample whereas the bottom panel shows the N-Loud sample.

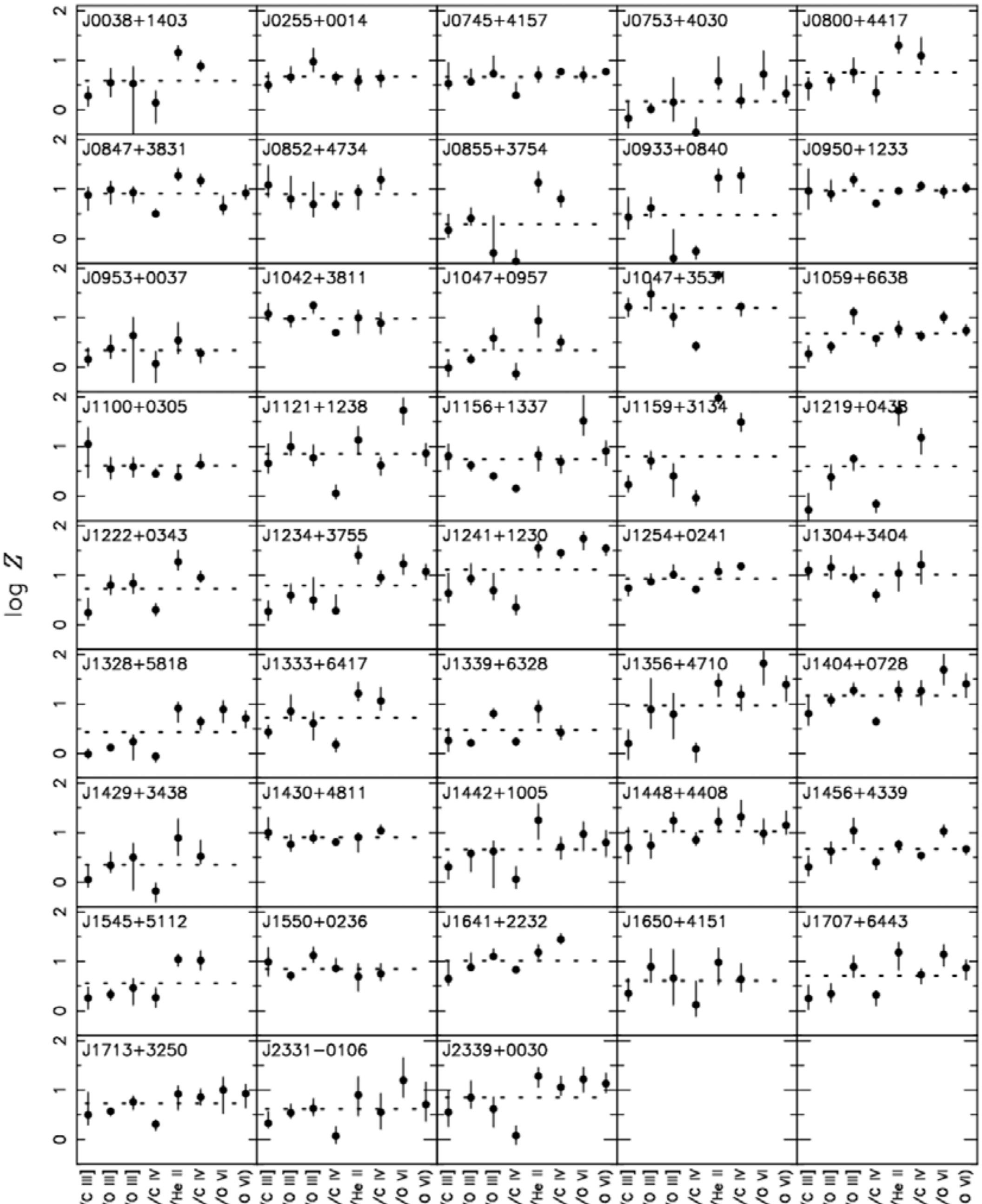

**Figure 5.** The metallicities determined from each of the indicated line-intensity ratios, and the mean metallicity (the horizontal dotted line in each panel), as initially measured. Note that line ratios used to find the individual $Z_{ind}$ occur in the same order within each column of panels, and also that not all objects have $Z_{ind}$ for the last two line ratios (which require the O VI line).

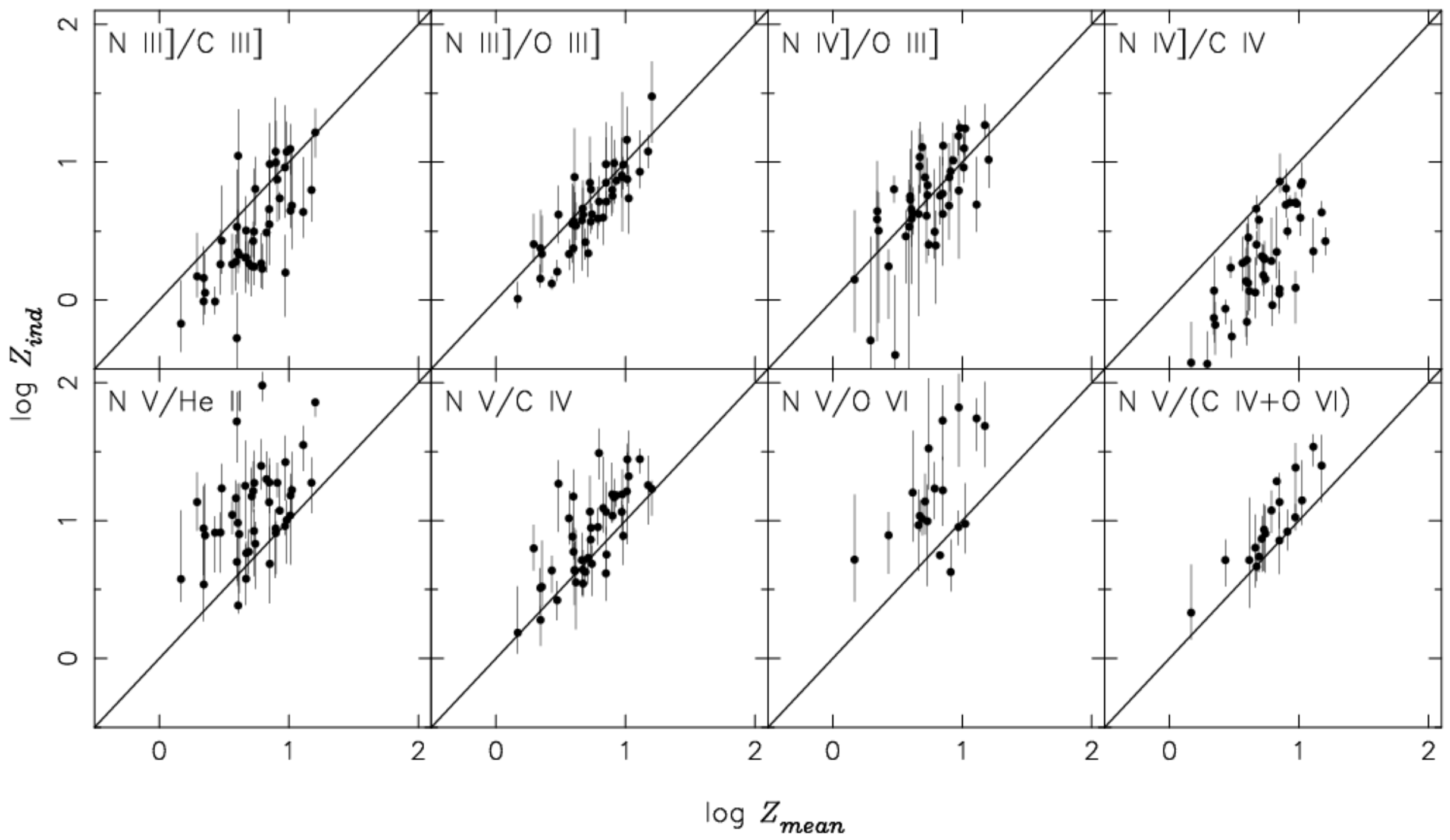

**Figure 6.** Metallicity from individual line-intensity ratios as a function of mean metallicity.

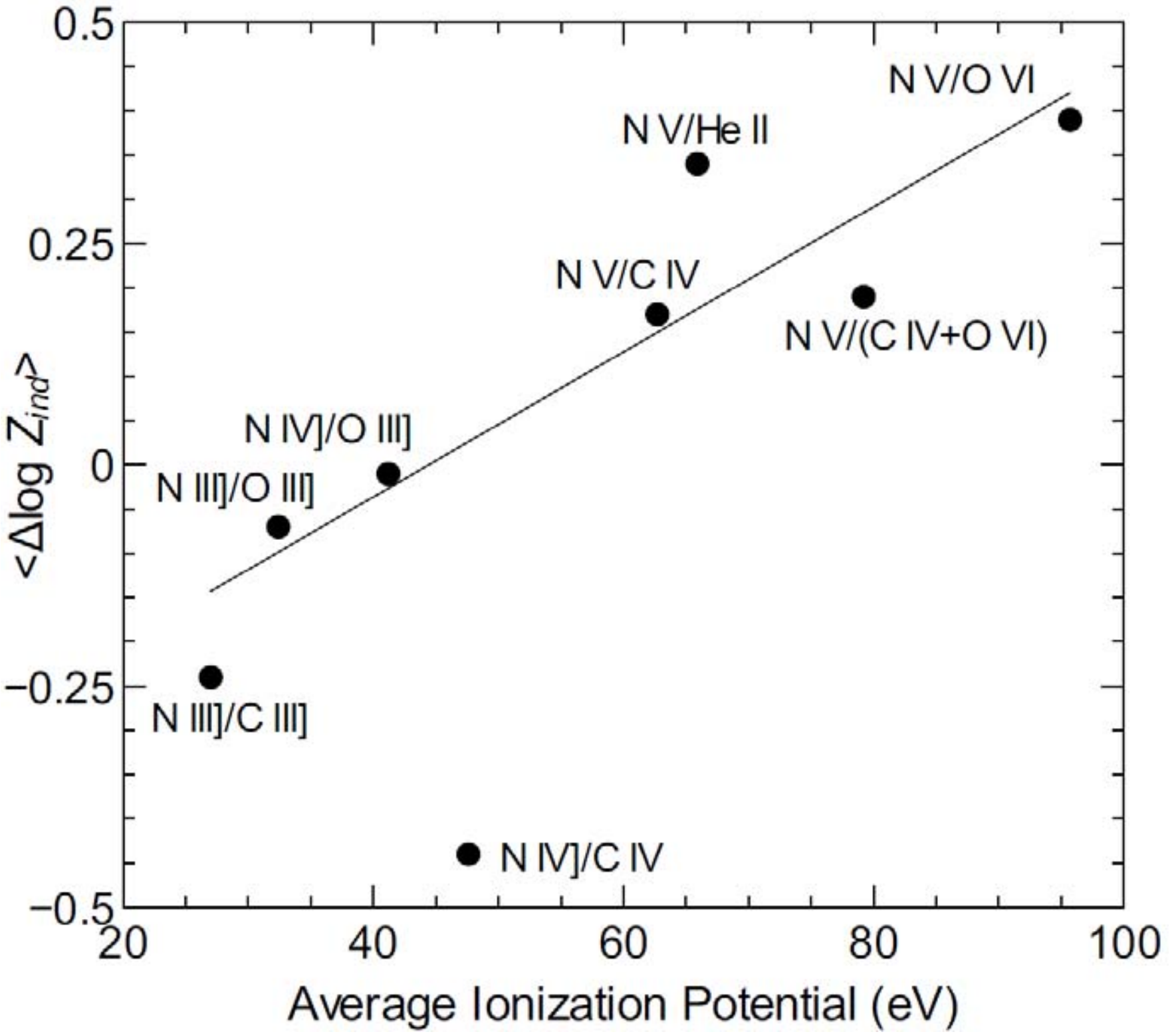

**Figure 7.** <Δlog $Z_{ind}$> vs. the average of the ionization potentials for creating the ions involved in each line ratio.

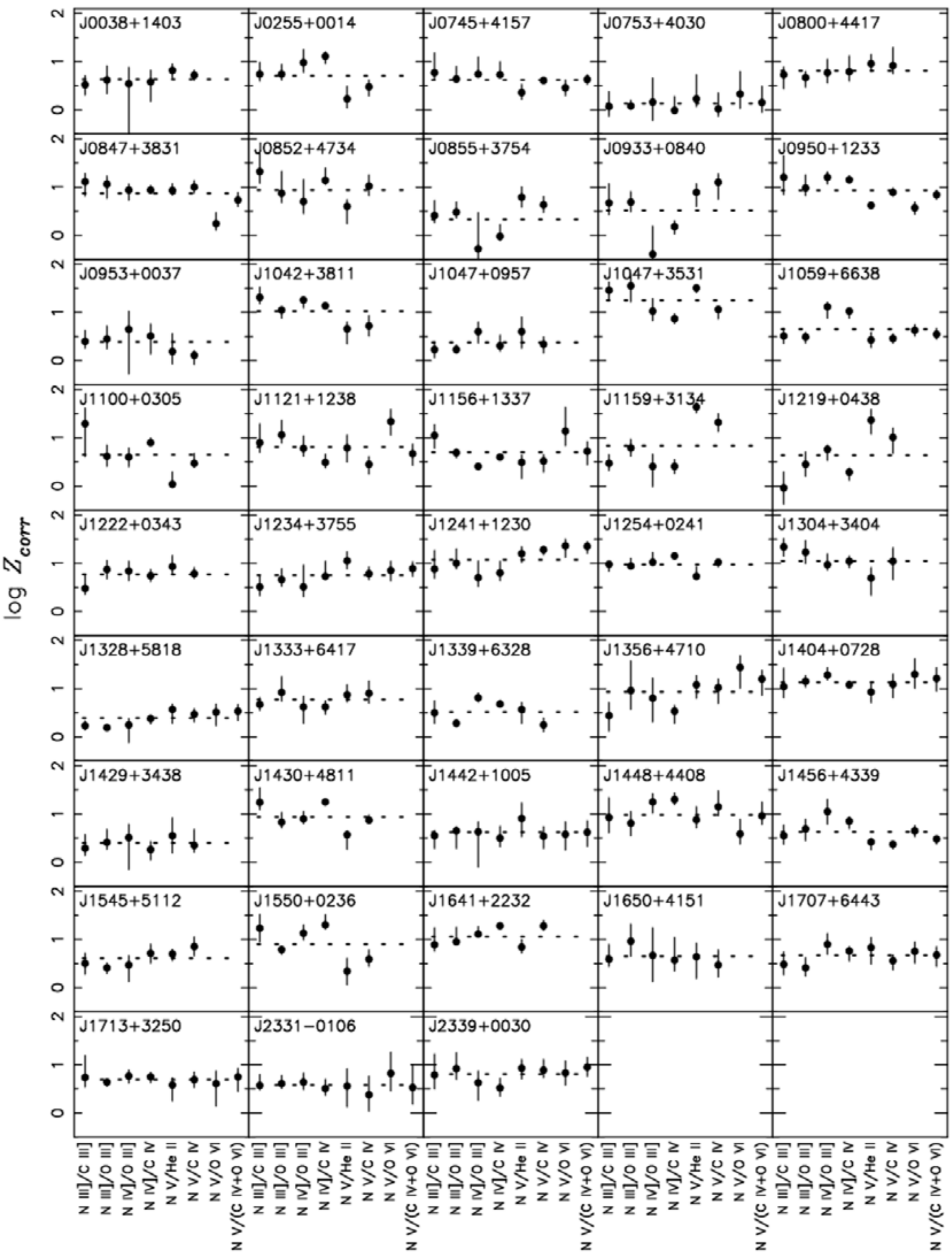

**Figure 8:** The metallicities determined from each of the indicated line ratios, and the mean metallicity (the horizontal dotted line in each panel), after correction for <Δlog $Z_{ind}$>.

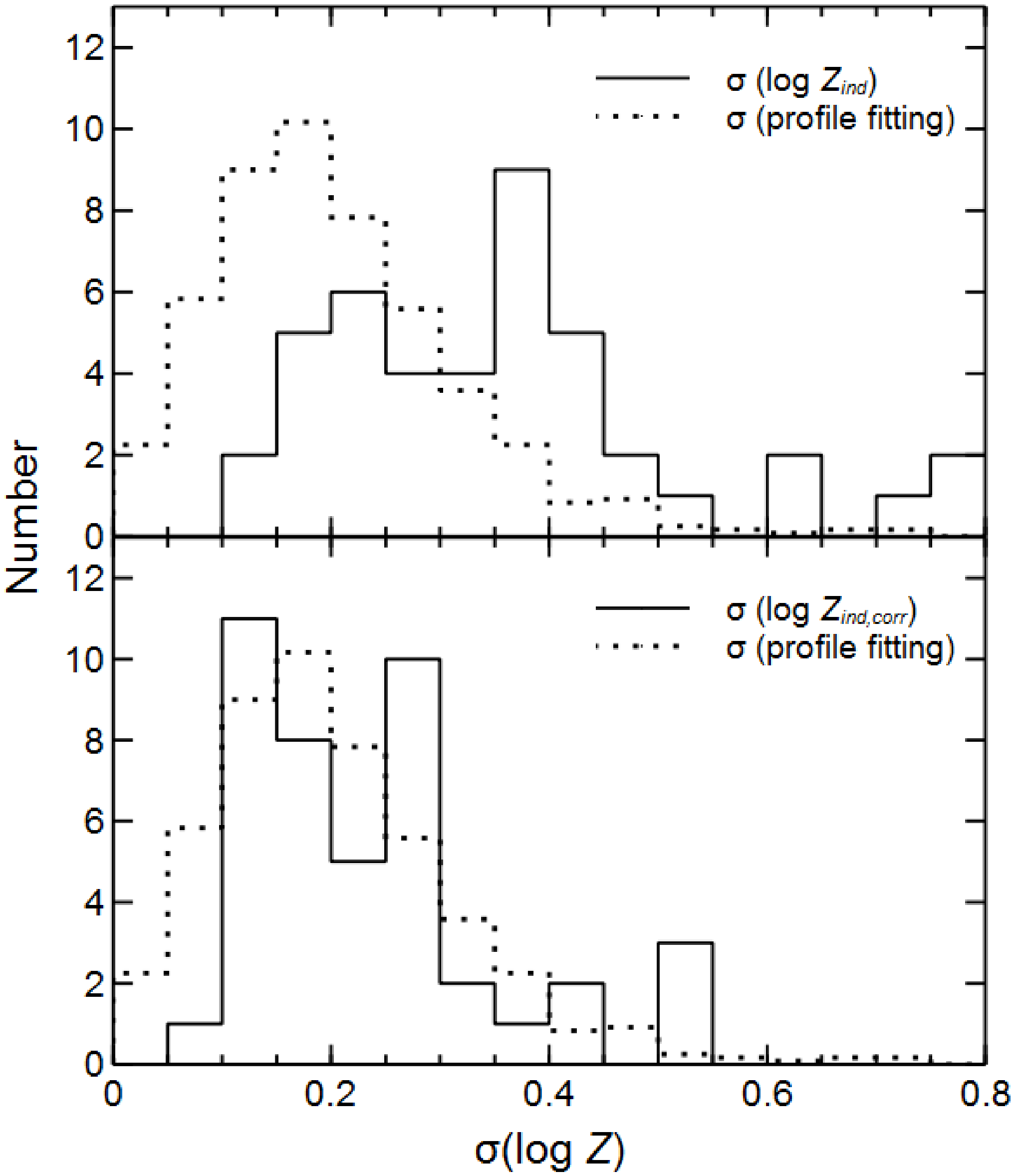

**Figure 9.** The distributions in σ(log *Z*), which is the scatter among the different $Z_{ind}$ measurements made for each individual QSO. The solid line in the upper panel is for the N-loud sample before applying the corrections for <Δlog $Z_{ind}$>, and the solid line in the lower panel is after applying those corrections. In both panels, the dotted line shows the distribution of uncertainties in the individual $Z_{ind}$ measurements due to the profile fitting uncertainties, for all measured $Z_{ind}$ for all QSOs in the sample.

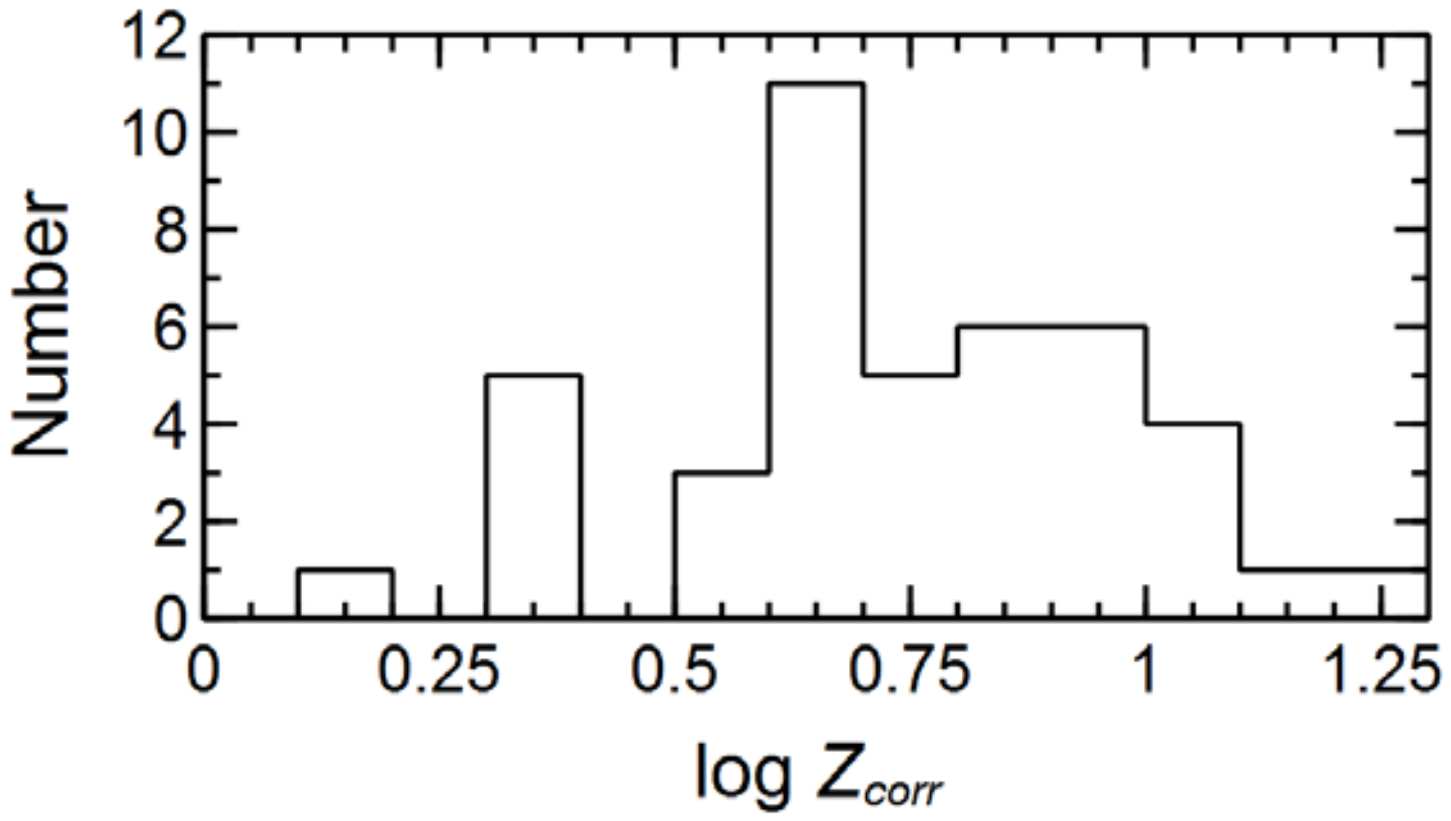

**Figure 10**. The distribution of log $Z_{corr}$ values for the N-Loud sample.

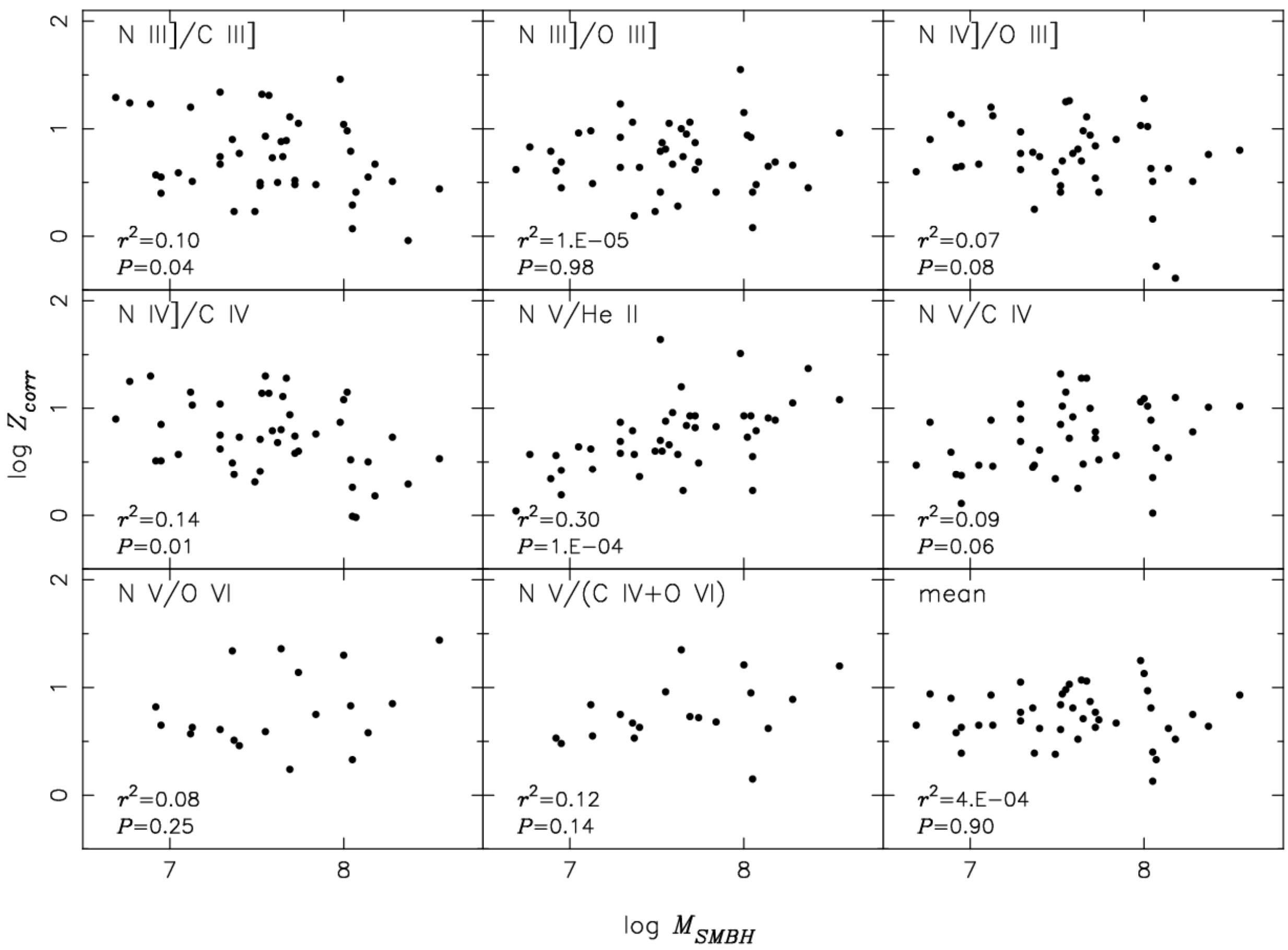

**Figure 11**: Metallicity from individual line-intensity ratios and their mean (bottom-right panel), as a function of supermassive black hole mass ($M_{SMBH}$). Each panel shows the coefficient of determination $r^2$ and the probability $P$ of obtaining an equally good correlation by chance.